\documentclass[fleqn,usenatbib]{mnras}

\usepackage{newtxtext,newtxmath}
\usepackage[T1]{fontenc}

\usepackage{graphicx}	
\usepackage{amsmath}	
\usepackage[dvipsnames]{xcolor}
\usepackage[flushleft]{threeparttable}
\usepackage{verbatim}
\usepackage{array}
\usepackage{tabularx}
\usepackage{colortbl}
\usepackage{hyperref}
\usepackage{float}
\usepackage{subcaption}
\usepackage{booktabs}
\usepackage{adjustbox}
\usepackage{pdflscape}
\usepackage{booktabs}
\usepackage{inputenc}
\usepackage{longtable}
\usepackage{xcolor, soul}
\sethlcolor{yellow}

\DeclareRobustCommand{\VAN}[3]{#2}
\let\VANthebibliography\thebibliography
\def\thebibliography{\DeclareRobustCommand{\VAN}[3]{##3}\VANthebibliography}

\newcommand{\kepler}{{\it Kepler}}

\newcommand{\TESS}{{\it TESS}}
\newcommand{\tess}{{\it TESS}}

\newcommand{\gaia}{{\it Gaia}}

\newcommand{\twomass}{{\it 2MASS}}

\newcommand{\jwst}{{\it JWST}}

\newcommand{\HARPS}{{\it HARPS}}
\newcommand{\harps}{{\it HARPS}}

\newcommand{\espresso}{{\it ESPRESSO}}

\newcommand{\lco}{{\it LCOGT}}
\newcommand{\lcogt}{{\it LCOGT}}
\newcommand{\LCO}{{\it LCOGT}}

\newcommand{\SOAR}{{\it SOAR}}
\newcommand{\soar}{{\it SOAR}}

\newcommand{\apass}{{\it APASS}}

\newcommand{\kms}{km\,s$^{-1}$}
\newcommand{\ms}{m\,s$^{-1}$}

\newcommand{\tc}{\mbox{$T_{\rm c}$}}

\newcommand{\rpl}{\mbox{$R_{\rm p}$}}
\newcommand{\mstar}{\mbox{$M_{\rm *}$}}
\newcommand{\rstar}{\mbox{$R_{\rm *}$}}

\newcommand{\msun}{\mbox{$M_{\odot}$}}
\newcommand{\rsun}{\mbox{$R_{\odot}$}}

\newcommand{\fsun}{\mbox{$S_{\odot}$}}
\newcommand{\rearth}{$R_{\oplus}$}
\newcommand{\mearth}{$M_{\oplus}$}
\newcommand{\gccc}{g\,cm$^{-3}$}

\newcommand{\teff}{$T_{\rm eff}$}
\newcommand{\teq}{$T_{\rm eq}$}
\newcommand{\logg}{$\log g$}
\newcommand{\vturb}{$v_{\rm turb}$}

\newcommand{\vsini}{$v \sin i_*$}

\newcommand{\feh}{[Fe/H]}
\newcommand{\mgh}{[Mg/H]}
\newcommand{\sih}{[Si/H]}
\newcommand{\alh}{[Al/H]}
\newcommand{\tih}{[Ti/H]}
\newcommand{\ch}{[C/H]}
\newcommand{\oh}{[O/H]}
\newcommand{\cuh}{[Cu/H]}
\newcommand{\znh}{[Zn/H]}
\newcommand{\srh}{[Sr/H]}
\newcommand{\yh}{[Y/H]}
\newcommand{\zrh}{[Zr/H]}
\newcommand{\bah}{[Ba/H]}
\newcommand{\ceh}{[Ce/H]}
\newcommand{\ndh}{[Nd/H]}
\newcommand{\nih}{[Ni/H]}
\newcommand{\nah}{[Na/H]}

\newcommand{\TStar}{TOI-908} 
\newcommand{\Tstar}{TOI-908} 
\newcommand{\TTIC}{TIC-350153977} 
\newcommand{\TTwomass}{J03323821-81150267}
\newcommand{\TGaiaId}{46192380870592067842} 

\newcommand{\TRa}{\mbox{$03^{\rmn{h}} 32^{\rmn{m}} 38\fs26$}} 
\newcommand{\TDec}{\mbox{$-81\degr 15\arcmin 02\farcs68$}} 
\newcommand{\TPropRa}{\mbox{$5.898\pm0.013$}} 
\newcommand{\TPropDec}{\mbox{$-0.353\pm0.015$}} 
\newcommand{\TPropTotal}{\mbox{$5.909\pm0.012$}} 
\newcommand{\TParallax}{\mbox{$5.686\pm0.010$}} 
\newcommand{\TDistance}{\mbox{$175.70^{+4.22}_{-0.59}$}} 
\newcommand{\GaiaRV}{\mbox{$9.071\pm0.347$}} 

\newcommand{\TTESSmag}{\mbox{$10.651\pm0.006$}} 
\newcommand{\TTESSmagshort}{\mbox{$10.7$}} 
\newcommand{\TBmag}{\mbox{$12.005\pm0.169$}} 
\newcommand{\TVmag}{\mbox{$11.316\pm0.012$}} 
\newcommand{\TGMag}{\mbox{$11.1061\pm0.0004$}} 
\newcommand{\TJmag}{\mbox{$10.04\pm0.02$}} 
\newcommand{\THmag}{\mbox{$9.734\pm0.026$}} 
\newcommand{\TKmag}{\mbox{$9.637\pm0.021$}} 
\newcommand{\TGaiaMagBP}{\mbox{$11.4711\pm0.0009$}} 
\newcommand{\TGaiaMagRP}{\mbox{$10.6024\pm0.8146$}} 

\newcommand{\Tloggporto}{\mbox{$4.45\pm0.03$}} 
\newcommand{\Teffporto}{\mbox{$5626\pm61$}} 
\newcommand{\Tstarvsiniporto}{\mbox{$2.560\pm0.636$}} 
\newcommand{\Tstarvturbporto}{\mbox{$0.913\pm0.022$}} 
\newcommand{\Tstarfehporto}{\mbox{$0.08\pm0.04$}} 
\newcommand{\Tstarmghporto}{\mbox{$0.10\pm0.03$}}
\newcommand{\Tstarsihporto}{\mbox{$0.07\pm0.02$}}
\newcommand{\Tstaralhporto}{\mbox{$0.08\pm0.04$}}
\newcommand{\Tstartihporto}{\mbox{$0.09\pm0.03$}}
\newcommand{\Tstarageporto}{\mbox{$4.6\pm1.5$}}
\newcommand{\Tstarohporto}{\mbox{$0.06\pm0.13$}} 
\newcommand{\Tstarchporto}{\mbox{$0.00\pm0.03$}}
\newcommand{\Tstarcuhporto}{\mbox{$0.08\pm0.03$}}
\newcommand{\Tstarznhporto}{\mbox{$0.05\pm0.02$}}
\newcommand{\Tstarsrhporto}{\mbox{$0.16\pm0.08$}}
\newcommand{\Tstaryhporto}{\mbox{$0.13\pm0.07$}}
\newcommand{\Tstarzrhporto}{\mbox{$0.12\pm0.03$}}
\newcommand{\Tstarbahporto}{\mbox{$0.07\pm0.04$}}
\newcommand{\Tstarcehporto}{\mbox{$0.08\pm0.02$}}
\newcommand{\Tstarndhporto}{\mbox{$0.12\pm0.03$}}
\newcommand{\Tstarnihporto}{\mbox{$0.07\pm0.02$}}
\newcommand{\Tstarnahporto}{\mbox{$0.10\pm0.03$}}

\newcommand{\Tstarmassexo}{\mbox{$0.950\pm0.010$}} 
\newcommand{\Tstarmassexoshort}{\mbox{$0.95\pm0.01$}} 
\newcommand{\Tstarradiusexo}{\mbox{$1.028\pm0.030$}} 
\newcommand{\Tstarradiusexoshort}{\mbox{$1.03\pm0.03$}} 
\newcommand{\Tstardensityexo}{\mbox{$1.235^{+0.091}_{-0.102}$}}
\newcommand{\Tstarperiodexo}{\mbox{$21.932\pm6.167$}} 

\newcommand{\ldu}{\mbox{$0.296\pm0.242$}}
\newcommand{\ldv}{\mbox{$0.164\pm0.291$}}

\newcommand{\Tplanetb}{TOI-908\,b}
\newcommand{\Tperiodb}{\mbox{$3.183792\pm0.000007$}} 
\newcommand{\Tperiodbshort}{\mbox{$3.18$}} 
\newcommand{\TTDurfullb}{\mbox{$2.457^{+0.120}_{-0.102}$}} 
\newcommand{\TTDurcutb}{\mbox{$2.268^{+2.225}_{-1.437}$}} 
\newcommand{\Tcb}{\mbox{$2384.292\pm0.002$}} 
\newcommand{\TMassb}{\mbox{$16.137^{+4.112}_{-4.039}$}} 
\newcommand{\TMassbshort}{\mbox{$16.1\pm4.1$}} 
\newcommand{\TRadiusb}{\mbox{$3.186\pm0.155$}} 
\newcommand{\TRadiusbshort}{\mbox{$3.18\pm0.16$}} 
\newcommand{\Tdensityb}{\mbox{$2.742^{+0.241}_{-0.353}$}} 
\newcommand{\Tdensitybshort}{\mbox{$2.7^{+0.2}_{-0.4}$}} 
\newcommand{\Trorb}{\mbox{$0.028\pm0.002$}} 
\newcommand{\Timpactb}{\mbox{$0.536\pm0.191$}} 
\newcommand{\Tkb}{\mbox{$7.244\pm1.768$}} 
\newcommand{\Tincb}{\mbox{$86.475^{+1.258}_{-1.260}$}} 
\newcommand{\Taub}{\mbox{$0.041657\pm0.000002$}} 
\newcommand{\Teqb}{\mbox{$1317\pm38$}}
\newcommand{\Teccb}{\mbox{$0.132\pm0.091$}}
\newcommand{\Tomegab}{\mbox{$35.856\pm94.972$}}
\newcommand{\TSMb}{\mbox{$76\pm6.7$}}

\newcommand{\Tfluxb}{\mbox{$80.884\pm0.006$}}

\title{TOI-908: a planet at the edge of the Neptune desert transiting a G-type star}

\author[F. Hawthorn et al.]{\parbox{\textwidth}{\Large
Faith~Hawthorn$^{1,2}$,
Daniel~Bayliss$^{1,2}$,
David~J.~Armstrong$^{1,2}$,
Jorge~Fern\'andez~Fern\'andez$^{1,2}$,
Ares~Osborn$^{1,2}$,
S\'ergio~G.~Sousa$^{3,4}$,
Vardan~Adibekyan$^{3,4}$,
Jeanne~Davoult$^{5}$,
Karen~A.~Collins$^{6}$,
Yann~Alibert$^{5}$,
Susana~C.~C.~Barros$^{3,4}$,
Fran\c{c}ois~Bouchy$^{7}$,
Matteo~Brogi$^{1,2,8}$,
David~R.~Ciardi$^{9}$,
Tansu~Daylan$^{10}$,
Elisa~Delgado~Mena$^{3}$,
Olivier~D.~S.~Demangeon$^{3,4}$,
Rodrigo~F.~D\'iaz$^{11}$,
Tianjun~Gan$^{12}$,
Keith~Horne$^{13}$,
Sergio~Hoyer$^{14}$,
Jon~M.~Jenkins$^{15}$,
Eric~L.~N.~Jensen$^{16}$,
John~F.~Kielkopf$^{17}$,
Veselin~Kostov$^{18}$,
David~W.~Latham$^{6}$,
Alan~M.~Levine$^{19}$,
Jorge~Lillo-Box$^{20}$,
Louise~D.~Nielsen$^{21}$,
Hugh~P.~Osborn$^{5,19}$,
George~R.~Ricker$^{19}$,
Jos\'e~Rodrigues$^{3,4}$,
Nuno~C.~Santos$^{3,4}$,
Richard~P.~Schwarz$^{6}$,
Sara~Seager$^{19,22,23}$,
Juan~Serrano~Bell$^{24}$,
Avi~Shporer$^{19}$,
Chris~Stockdale$^{25}$,
Paul~A.~Str\o{}m$^{1,2}$,
Peter~Tenenbaum$^{15,26}$,
St\'ephane~Udry$^{7}$,
Peter~J.~Wheatley$^{1,2}$,
Joshua~N.~Winn$^{10}$,
Carl~Ziegler$^{27}$
}
\vspace{0.2cm}
\\
\parbox{\textwidth}{
The authors' affiliations are shown in Appendix \ref{sec:affiliations}.\\
*E-mail: faith.hawthorn@warwick.ac.uk}\vspace{-0.3cm}}

\date{Accepted XXX. Received YYY; in original form ZZZ}

\pubyear{2022}

\begin{document}
\label{firstpage}
\pagerange{\pageref{firstpage}--\pageref{lastpage}}
\maketitle


\begin{abstract}
We present the discovery of an exoplanet transiting \Tstar\ (\TTIC) using data from \TESS\ sectors 1, 12, 13, 27, 28 and 39.  TOI-908 is a $T=$ \TTESSmagshort\,mag G-dwarf (\teff\,=\,\Teffporto\,K) solar-like star with a mass of \Tstarmassexo\,\msun\, and a radius of \Tstarradiusexo\,\rsun. The planet, \Tplanetb, is a \TRadiusbshort\,\rearth\, planet in a \Tperiodbshort\,day orbit.  Radial velocity measurements from \harps\ reveal \Tplanetb\ has a mass of approximately \TMassbshort\,\mearth, resulting in a bulk planetary density of \Tdensitybshort\,\gccc. \Tplanetb\ lies in a sparsely-populated region of parameter space known as the Neptune desert. The planet likely began its life as a sub-Saturn planet before it experienced significant photoevaporation due to X-rays and extreme ultraviolet radiation from its host star, and is likely to continue evaporating, losing a significant fraction of its residual envelope mass.

\end{abstract}



\section{Introduction} \label{sec:intro}
In the years since \kepler\ \citep{kepler2010} and during the lifetime of the \tess\ mission \citep{Ricker2015}, a distinct dearth of exoplanets between $3 < R_\oplus < 4$ with orbital periods less than $\sim5$\,days has been discovered, the so-called `Neptune desert', `evaporation desert' or `sub-Jovian desert' \citep{szabokiss2011, owenlai2018, mazeh2016}. Several theories have been made to explain this sparse parameter space - including that these relatively low-mass planets have had their gaseous H/He envelopes stripped away by high levels of irradiation from their host stars \citep{OwenWu17}, leaving behind a dense core (eg. TOI-849\,b; \citealt{849b}), whilst some are still undergoing this process (eg. LTT 9779\,b, \citealt{jenkins2020}; NGTS-4\,b, \citealt{west2019}; TOI-969\,b, \citealt{lillobox2022}).

Many of these planets are readily observable with missions such as \tess\ and \jwst, and also with ground-based spectroscopic instruments such as \harps\ \citep{HARPS} and \espresso\ \citep{ESPRESSO} due to their close orbits and short periods. The \harps-NOMADS program (PI Armstrong, 1108.C-0697) aims to significantly increase the number of planet confirmations in the Neptune desert with precise masses and radii. In this paper we present one such detection of \Tplanetb, a sub-Neptune transiting a G-type star. Precise measurements of these parameters is highly important in allowing us to constrain the density and internal structure of these planets, assisting in our understanding of the formation and evolution mechanisms that place these planets in the desert.

This paper is structured as follows: we present our observations of \Tstar\ from \tess\ and \lcogt\ photometry, \harps\ spectroscopy and \soar-HRCam imaging in Section~\ref{sec:obs}, our stellar analysis and global joint modelling of the data in Section~\ref{sec:methods}, and our results and discussion of our findings in Section~\ref{sec:results}, including the position of the planet in the Neptune desert and the evolution of its envelope. We finally present our conclusions in Section~\ref{sec:conc}.

\section{Observations} \label{sec:obs}

\subsection{TESS photometry}
\label{sec:tessphot}

\begin{figure}
    \centering
    \includegraphics[width=0.5\textwidth]{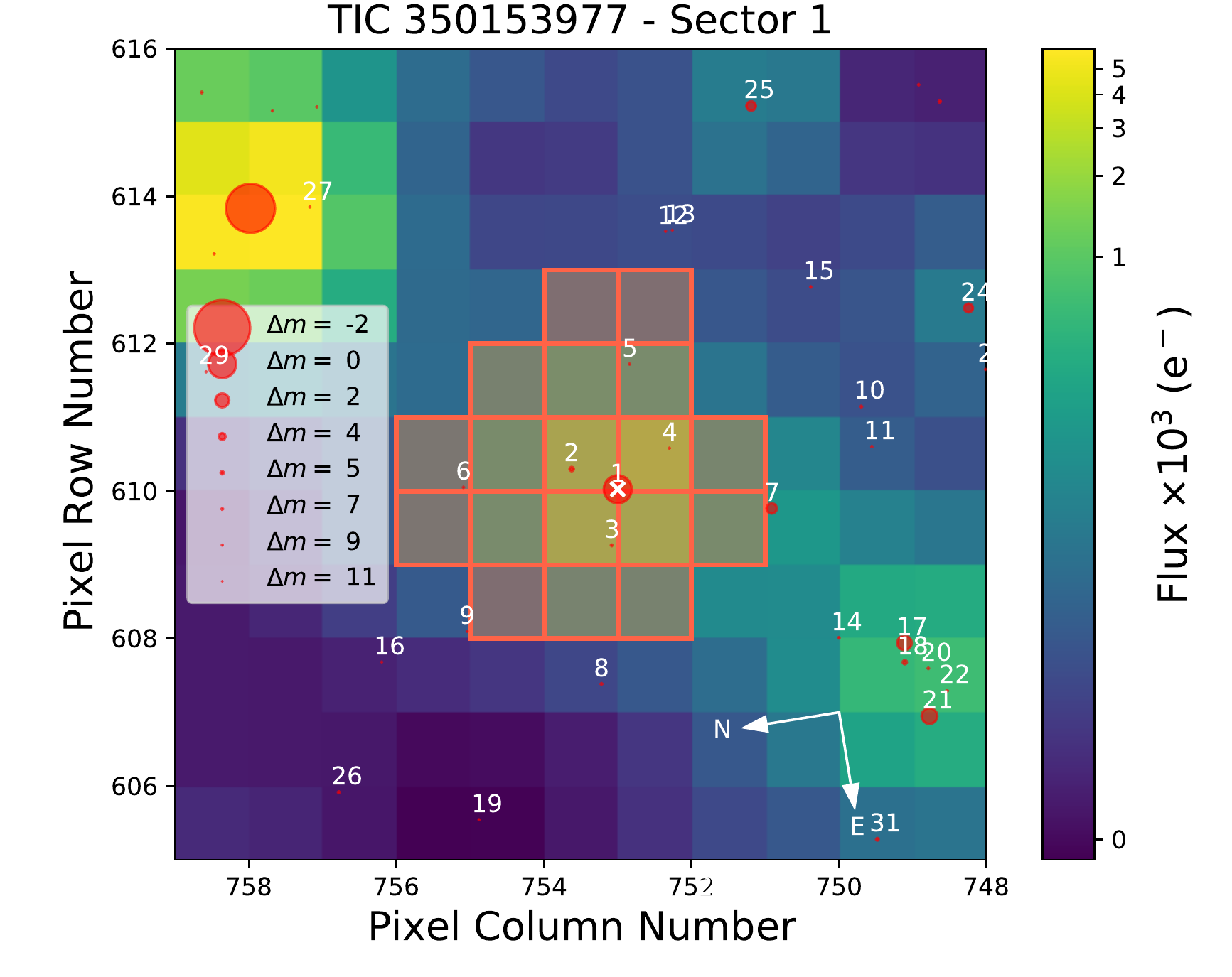}
    \caption{Target Pixel File (TPF) from \tess\ sector 1 with \Tstar\ marked with a white cross. Other sources from \gaia\ DR2 are marked with red circles sized by scaled magnitudes relative to the target, ranked by distance. The aperture mask is indicated by the red outline.}
    \label{fig:tpfgaia}
\end{figure}

\TESS\ is a space-based NASA telescope that is currently performing a survey search for transiting exoplanets around bright host stars. It is equipped with four cameras for a total combined FOV (Field-Of-View) of 24 $\times$ 96\textdegree. It splits the sky into 13 sectors per hemisphere, each of which is observed for approximately 27\,days, making \TESS\ a key mission in detecting short-period transiting exoplanets. \TESS\ observed the bright star \Tstar\ (\TTIC) in sectors 1, 12 and 13 during Cycle 1 of operation (2018-07-25 to 2019-07-17) at a cadence of 30\,minutes, and sectors 27, 28 and 39 during Cycle 3 (2020-07-05 to 2021-06-24) at a cadence of 10\,minutes. The target location at a declination of approximately -81$^{\circ}$ means it lies close to the \tess\ Continuous Viewing Zone (CVZ), and therefore also close to the CVZ of \jwst. \Tstar\ is a $T=$\,10.7~mag G-dwarf with an effective temperature of \Teffporto\,K (see Section~\ref{sec:stellaranalysis}). Details of \Tstar\ including identifiers, astrometric and photometric properties are presented in Table~\ref{tab:star_props}, and a full list of the \TESS\ sector details are set out in Table~\ref{tab:photobs}. We present the Target Pixel File (TPF; created with \texttt{tpfplotter}\footnote{\url{https://github.com/jlillo/tpfplotter}} from \citealt{tpfplotter}) in Figure~\ref{fig:tpfgaia} with \Tstar\ as the central object with \gaia\ DR2 sources, scaled magnitudes for each object ranked by distance from \Tstar\ and the aperture mask used for photometry extraction with a TIC contamination ratio of 0.040531.

\begin{figure*}
    \centering
    \begin{subfigure}[b]{0.95\textwidth}
       \includegraphics[width=1\linewidth]{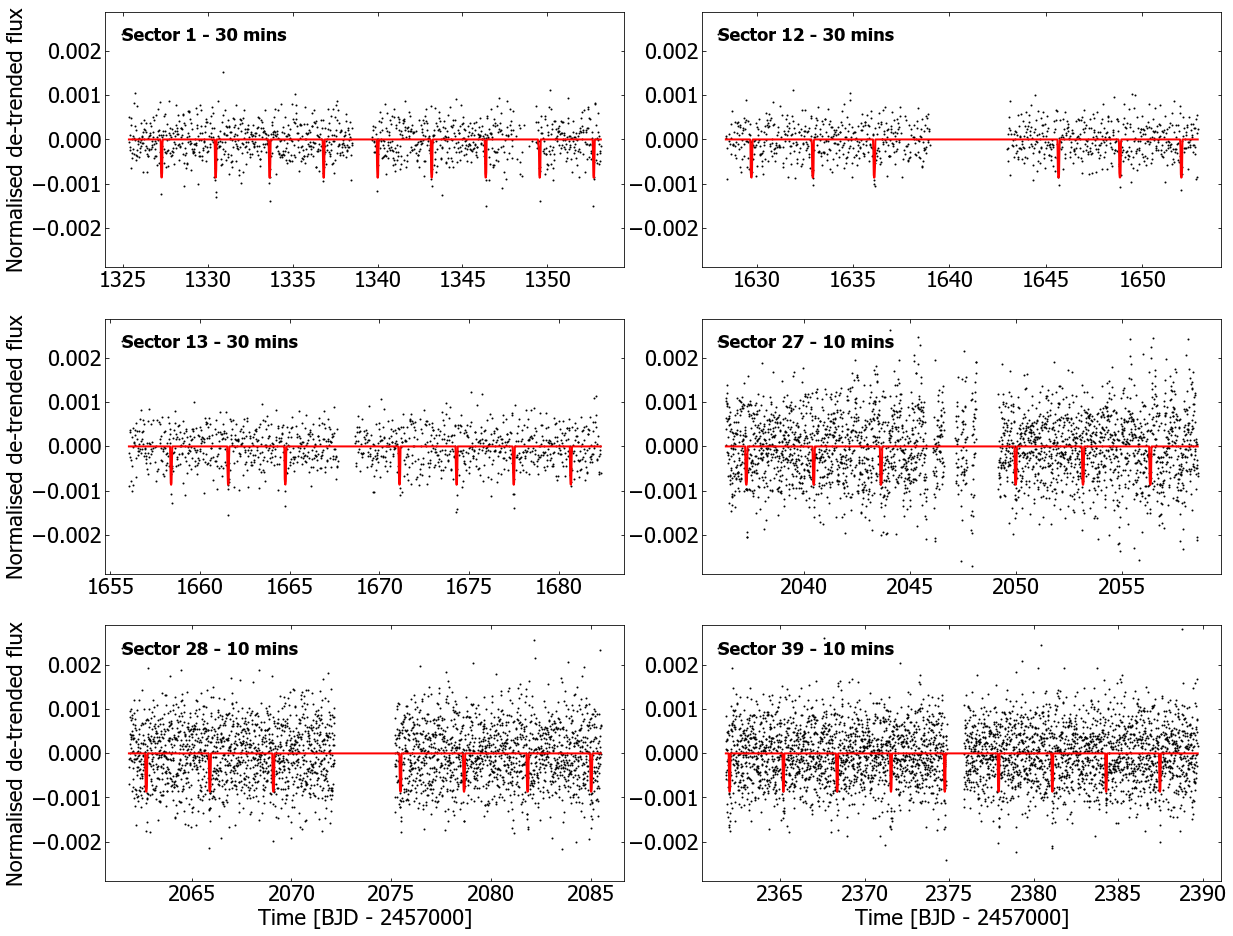}
    \end{subfigure}

    \begin{subfigure}[b]{0.8\textwidth}
       \includegraphics[width=1\linewidth]{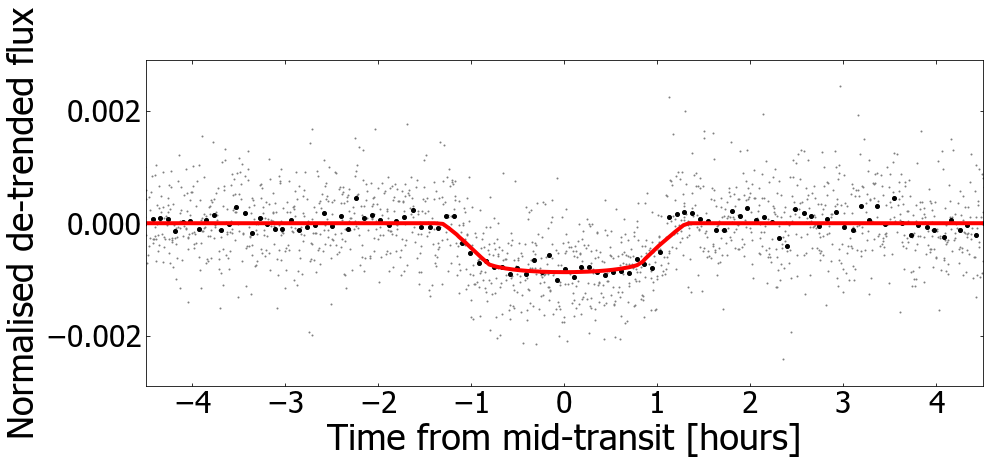}
    \end{subfigure}

\caption{\textbf{Top panels:} \tess\ PDCSAP lightcurves from labelled sectors at labelled cadences with the GP model removed, overplotted with the best-fitting transit models for \Tplanetb\ in red. \textbf{Bottom panel:} \tess\ PDCSAP lightcurves from all sectors (grey points), phase folded to a period corresponding to that of \Tplanetb\ overplotted with the best-fitting transit model in red and binned to 10 minute intervals (black points).}
\label{fig:tessall} 
\end{figure*}

The candidate was alerted as a TOI (\TESS\ Object of Interest, \citealt{toi2021}) and designated \Tstar.01 (hereafter \Tplanetb), based on the identification of a \Tperiodbshort\,day transit signal from the SPOC pipeline \citep{jenkins2002,jenkins2010, jenkins2020}. This pipeline uses the PDCSAP \citep[Presearch Data Conditioning Simple Aperture Photometry;][]{Stumpe2012, Stumpe2014, Smith2012} light-curves from the \tess\ HLSP (High Level Science Products), which removes instrumental and some stellar trends from the SAP (Simple Aperture Photometry) data, but retains local features such as transits. The transits of \Tplanetb\ passed the diagnostic tests after fitting with a limb-darkened transit model \citep{li2019} and were reported in the Data Validation Report \citep{twicken2018}. The PDCSAP light-curves are used in the joint model described in Section~\ref{sec:jointmodel}. We present the normalised light curves from each Sector of \tess\ along with the generated best-fitting transit models for \Tplanetb\ in Figure~\ref{fig:tessall}.

\begin{center}
\begin{table}
    \centering
    \caption{Stellar parameters of \TStar.}
    \label{tab:star_props}
    \begin{threeparttable}
    \begin{tabularx}{0.95\columnwidth}{ l  p{0.33\linewidth} X }
    \toprule
    \textbf{Property} & \textbf{Value} & \textbf{Source} \\
    \hline
    \textbf{Identifiers}                & & \\
    TIC ID      & \href{https://exofop.ipac.caltech.edu/tess/target.php?id=350153977}{\TTIC}              & TICv8 \\
    2MASS ID    & \TTwomass      & \twomass \\
    Gaia ID     & \TGaiaId   & \gaia\ DR3 \\
    \hline
    \textbf{Astrometric properties}     & & \\
    R.A. (J2015.5)  & \TRa           & \gaia\ DR3 \\
    Dec (J2015.5)   & \TDec          & \gaia\ DR3 \\
    Parallax (mas)  & \TParallax      & \gaia\ DR3 \\
    Distance (pc)   & \TDistance      & \\
    $\mu_{\rm{R.A.}}$ (mas yr$^{-1}$)       & \TPropRa   & \gaia\ DR3 \\
    $\mu_{\rm{Dec}}$ (mas yr$^{-1}$)        & \TPropDec  & \gaia\ DR3 \\
    $\mu_{\rm{Total}}$ (mas yr$^{-1}$)      & \TPropTotal  & \gaia\ DR3 \\
    RV\textsubscript{sys} (\kms) &\GaiaRV & \gaia\ DR3 \\
    \gaia\ non-single star flag & 0* & \gaia\ DR3 \\
    \hline
    \textbf{Photometric properties} & & \\
    \tess\ (mag)  & \TTESSmag    & TICv8 \\
    B (mag)     & \TBmag     & \apass \\
    V (mag)     & \TVmag      & \apass \\
    G (mag)     & \TGMag  & \gaia\ DR3 \\
    J (mag)     & \TJmag      & TICv8 \\
    H (mag)     & \THmag      & TICv8 \\
    K (mag)     & \TKmag      & TICv8 \\
    \gaia\ BP (mag)     & \TGaiaMagBP  & \gaia\ DR2 \\
    \gaia\ RP (mag)   & \TGaiaMagRP   & \gaia\ DR2 \\
    
    \bottomrule
    \end{tabularx}
    \begin{tablenotes}
    \item Sources: TICv8 \citep{Stassun2019}, \twomass\ \citep{Skrutskie2006}, \gaia\ Data Release 3 \citep{GAIA_DR3}, \apass\ \citep{apass}, \gaia\ Data Release 2 \citep{GAIA_DR2}
    \item * Indicates that the star does not belong to an astrometric, spectroscopic or eclipsing binary.
    \end{tablenotes}
    \end{threeparttable}
\end{table}
\end{center}


\begin{center}
\begin{table*}
    \centering
    \caption{Photometric observations of \TStar.}
    \label{tab:photobs}
    \begin{tabularx}{0.965\textwidth}{ l c c c c c c c}
    \toprule
    \textbf{Instrument} &\textbf{Aperture} &\textbf{Filter} &\textbf{Exposure time (s)} &\textbf{No. of images} &\textbf{UT night} &\textbf{Detrending} &\textbf{TTV (mins)}\\
    \hline
    \tess\ S01   & 0.105\,m   & \tess\textsuperscript{1}   & 1800   & 1337   & 2018 Jul 25 - 2018 Aug 22 & GP model & $-0.2^{+2.4}_{-2.2}$* \\
    \tess\ S12   & 0.105\,m   & \tess   & 1800   & 1340   & 2019 May 21 - 2019 Jun 18 & GP model & $5.2^{+2.8}_{-4.3}$\\
    \tess\ S13   & 0.105\,m   & \tess   & 1800   & 1365   & 2019 Jun 19 - 2019 Jul 17 & GP model & $-0.2^{+2.2}_{-2.6}$\\
    \lcogt-CTIO\textsuperscript{A} & 1.0\,m  & \textit{z\textsubscript{s}}\textsuperscript{2} & 35 & 302 & 2019 Sep 08 & Time, losses & $-12.3^{+4.2}_{-3.8}$ \\
    \lcogt-SAAO\textsuperscript{B}  & 1.0\,m    & \textit{z\textsubscript{s}}   & 35    & 252   & 2019 Dec 31 & Time, losses &  $-8.2^{+8.8}_{-6.7}$\\
    \tess\ S27   & 0.105\,m   & \tess   & 600   & 3508   & 2020 Jul 05 - 2020 Jul 30 & GP model & $-7.3^{+2.7}_{-3.2}$ \\
    \tess\ S28  & 0.105\,m   & \tess   & 600   & 3636   & 2020 Jul 31 - 2020 Aug 25 &  GP model & $4.7^{+4.3}_{-1.3}$\\
    \lcogt-SAAO & 1.0\,m    & \textit{z\textsubscript{s}}   & 35    & 227   & 2020 Sep 17 & Airmass & $-17.3^{+9.5}_{-4.5}$\\
    \lcogt-CTIO & 1.0\,m    & \textit{z\textsubscript{s}}   & 35    & 176   & 2020 Nov 30 & x-centroid & $-7.5^{+14.2}_{-7.8}$\\
    \lcogt-CTIO & 1.0\,m    & \textit{i\textsubscript{p}}\textsuperscript{3}   & 20    & 308   & 2020 Dec 16 & x-centroid, losses & $-5.1^{+8.8}_{-10.9}$\\
    \tess\ S39  & 0.105\,m   & \tess   & 120   & 4024   & 2021 May 27 & GP model & $0.9^{+1.3}_{-3.1}$\\
    \bottomrule
    \end{tabularx}
    \begin{tabularx}{0.87\textwidth}{ c p{0.07\textwidth} p{0.12\textwidth} }
        \textsuperscript{1}\tess\ custom, 600--1000 nm \hspace{3mm} \textsuperscript{2}PanSTARRS \textit{z}-short, $\lambda$\textsubscript{mid} = 8700\,nm, $\delta\lambda$ = 1040\,nm \hspace{3mm} \textsuperscript{3}SDSS \textit{i'}, $\lambda$\textsubscript{mid} = 7545\,nm, $\delta\lambda$ = 1290\,nm
    \end{tabularx}
    \begin{tabularx}{0.87\textwidth}{ c c}
        \textsuperscript{A}CTIO - Cerro Tololo Inter-American Observatory \hspace{3 mm} \textsuperscript{B}SAAO - South Africa Astronomical Observatory
    \end{tabularx}
    \begin{tabularx}{0.87\textwidth}{l}
    * Average TTV per \tess\ sector, calculated using \tc\,=\,\Tcb\,TBJD and $P$\,=\,\Tperiodb\,days \\ See Table~\ref{tab:ttvs} for full table of \tess\ TTVs.
    \end{tabularx}
\end{table*}
\end{center}


\subsection{LCOGT follow-up photometry}
\label{sec:lcophot}

\begin{figure}

\begin{subfigure}[b]{0.5\textwidth}
       \includegraphics[width=1\linewidth]{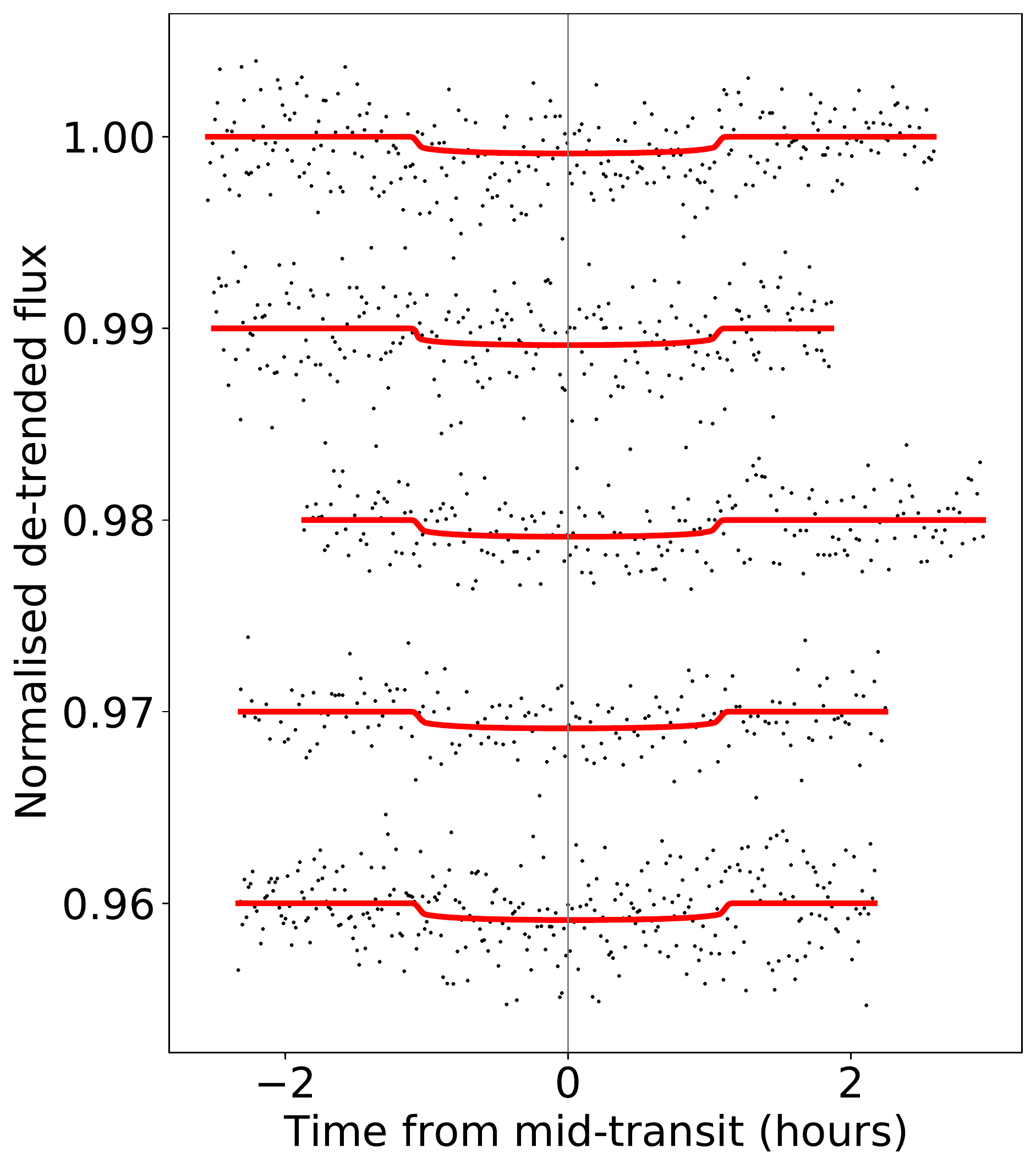}

    \end{subfigure}

    \begin{subfigure}[b]{0.5\textwidth}
       \includegraphics[width=1\linewidth]{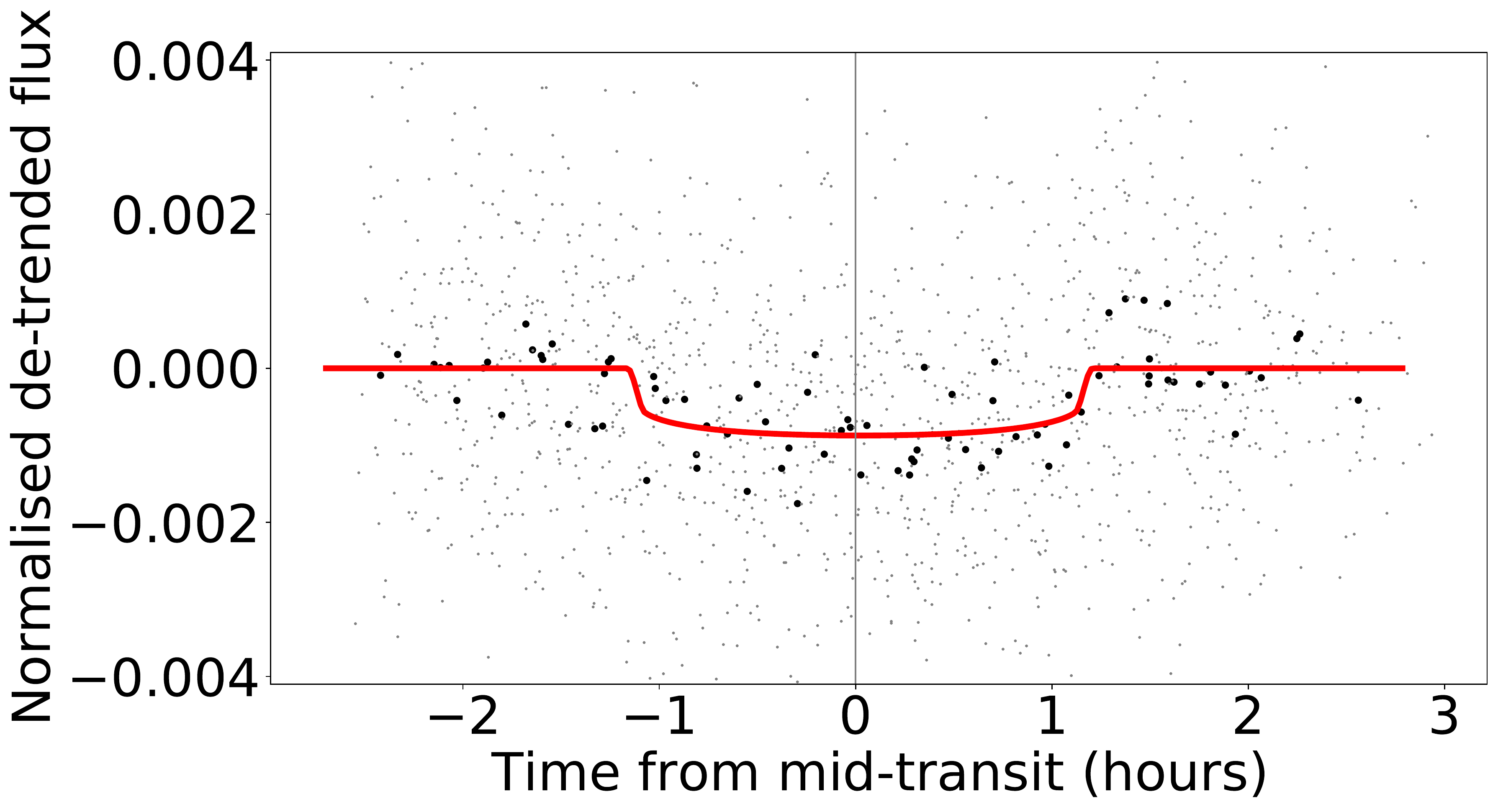}
    \end{subfigure}

\caption{\textbf{Top panel:} Photometric observations of \Tstar\ taken by the \LCO\ facilities as detailed in Table~\ref{tab:photobs} in order of observation date from top to bottom, overplotted with the best-fitting transit models in red, and offset vertically for clarity. \textbf{Bottom panel:} All \LCO\ lightcurves (grey points), phase folded to a period corresponding to that of \Tplanetb, overplotted with the best-fitting transit model in red and binned to 10 minute intervals (black points).}
\label{fig:lcoall} 
\end{figure}

The \lco\ \citep[Las Cumbres Observatory Global Telescope network;][]{lcogt2013} was used to take a total of five time-series transit photometry observations of \Tstar, each of which is detailed in Table~\ref{tab:photobs}. We used the {\tt TESS Transit Finder}, which is a customized version of the {\tt Tapir} software package \citep{Jensen:2013}, to schedule our transit observations. Each of the telescopes has a $26\arcmin\times26\arcmin$ FOV from 4096 $\times$ 4096 SINISTRO cameras with an image scale of 0.389\arcsec\,pix\textsuperscript{-1}, and observations were taken in the SDSS \textit{i'} and PanSTARRS \textit{z}-short bands. The \lcogt\ image data were processed using the standard \texttt{BANZAI} data reduction pipeline presented in \citet{McCully:2018}, and photometric data were extracted with the \texttt{AstroImageJ} analysis software detailed in \citet{Collins:2017} using circular photometric apertures with radii $7\farcs0$ or smaller, which exclude flux from the nearest known Gaia DR3 star $13\farcs5$ northwest of \Tstar. Parametric detrending vectors were selected by jointly fitting a transit model and linear combinations of zero, one, or two detrending parameters from the available detrending vectors airmass, time, sky-background, FWHM, x-centroid, y-centroid, and total comparison star counts (a proxy for atmospheric losses). The best zero, one, or two detrend vectors were retained if they improved the Bayesian information criterion (BIC) for the fit by at least a factor of two per detrend parameter. The detrending vectors selected for each light curve are shown in Table~\ref{tab:photobs}.  The \lco\ light-curves are used in the joint model in Section~\ref{sec:jointmodel}. A $\sim$900\,ppm transit-like event was detected in the follow-up light curves (see Section \ref{sec:results}), confirming that the \TESS-detected signal occurs on \Tstar\ relative to known \gaia\ DR3 stars. The \lcogt\ data is publicly available on the ExoFOP-\textit{TESS} website\footnote{\url{https://exofop.ipac.caltech.edu/tess/}}. We present the light curves and best-fitting transit models from \lcogt\ in Figure~\ref{fig:lcoall}.


\subsection{HARPS radial velocity observations}
\label{sec:harpsrv}

We obtained 42 spectra with the High Accuracy Radial velocity Planet Searcher \citep[\harps;][]{HARPS}. \harps\ is an echelle spectrograph mounted on the 3.6\,m ESO telescope at La Silla Observatory, Chile, capable of stabilised high-resolution measurements (R$\sim$115,000) at $\sim$1\,\ms\ precision \citep{HARPS}. These spectra were obtained through the \harps-NOMADS program (PI Armstrong, 1108.C-0697) from 2021-10-15 to 2022-01-26 in High Accuracy Mode (HAM) with a fibre diameter of 1\,\arcsec\ and an exposure time of 1800\,s, leading to a typical signal-to-noise of 40--50 per pixel at a wavelength of 550\,nm. The raw \harps\ data are processed using the standard data reduction pipeline presented in \citet{lovis2007} using the G2 spectral mask, which also includes measurements of the Full Width Half Maximum (FWHM), the line bisector span, the contrast of the Cross-Correlation Function (CCF) and the standard activity indicators of \textit{S}-index, H$\alpha$-index, Na-index and Ca-index. 

\begin{center}
\begin{table*}
    \centering
    \caption{\HARPS\ spectroscopic data for \Tstar. This table is available in its entirety online.}
    \label{tab:harpsobs}
    \begin{tabular}{c c c c c c c c c c c}  
    \toprule
    \textbf{Time (BJD} & \textbf{RV} & \textbf{RV error} & \textbf{FWHM} & \textbf{FWHM} & \textbf{Bisector} & \textbf{Bisector} & \textbf{Contrast} & \textbf{Contrast} &  \textbf{S-index\textsubscript{MW}} & \textbf{S-index\textsubscript{MW}} \\
    \textbf{-2457000)} & \textbf{(\ms)} & \textbf{(\ms)} & \textbf{(\ms)} & \textbf{error (\ms)} & \textbf{(\ms)} & \textbf{error (\ms)}  & & \textbf{error} & & \textbf{error} \\
    \hline
    2502.6901 & 9091.80 & 3.25 & 7074.23 & 10.00 & -21.19 & 4.60 & 49.379401 & 0.000001 & 0.15 & 0.01\\
    2503.7151 & 9106.48 & 2.92 & 7080.40 & 10.01 & -19.09 & 4.12 & 49.346258 & 0.000001 & 0.15 & 0.01\\
    2504.7080 & 9109.55 & 2.43 & 7067.72 & 10.00 & -20.01 & 3.43 & 49.306873 & 0.000001 & 0.16 & 0.01\\
    2505.6817 & 9096.90 & 2.92 & 7063.40 & 9.989 & -30.19 & 4.13 & 49.292743 & 0.000001 & 0.14 & 0.01 \\
    2505.7989 & 9095.50 & 2.50 & 7067.65 & 10.00 & -31.01 & 3.53 & 49.280238 & 0.000001 & 0.16 & 0.01\\
    ... & ...   & ...   & ...   & ...   & ...   & ... & ... & ... & ... & ... \\
\bottomrule
    \end{tabular}
\end{table*}
\end{center}

\begin{figure*}
    \centering
    \includegraphics[width=0.8\textwidth]{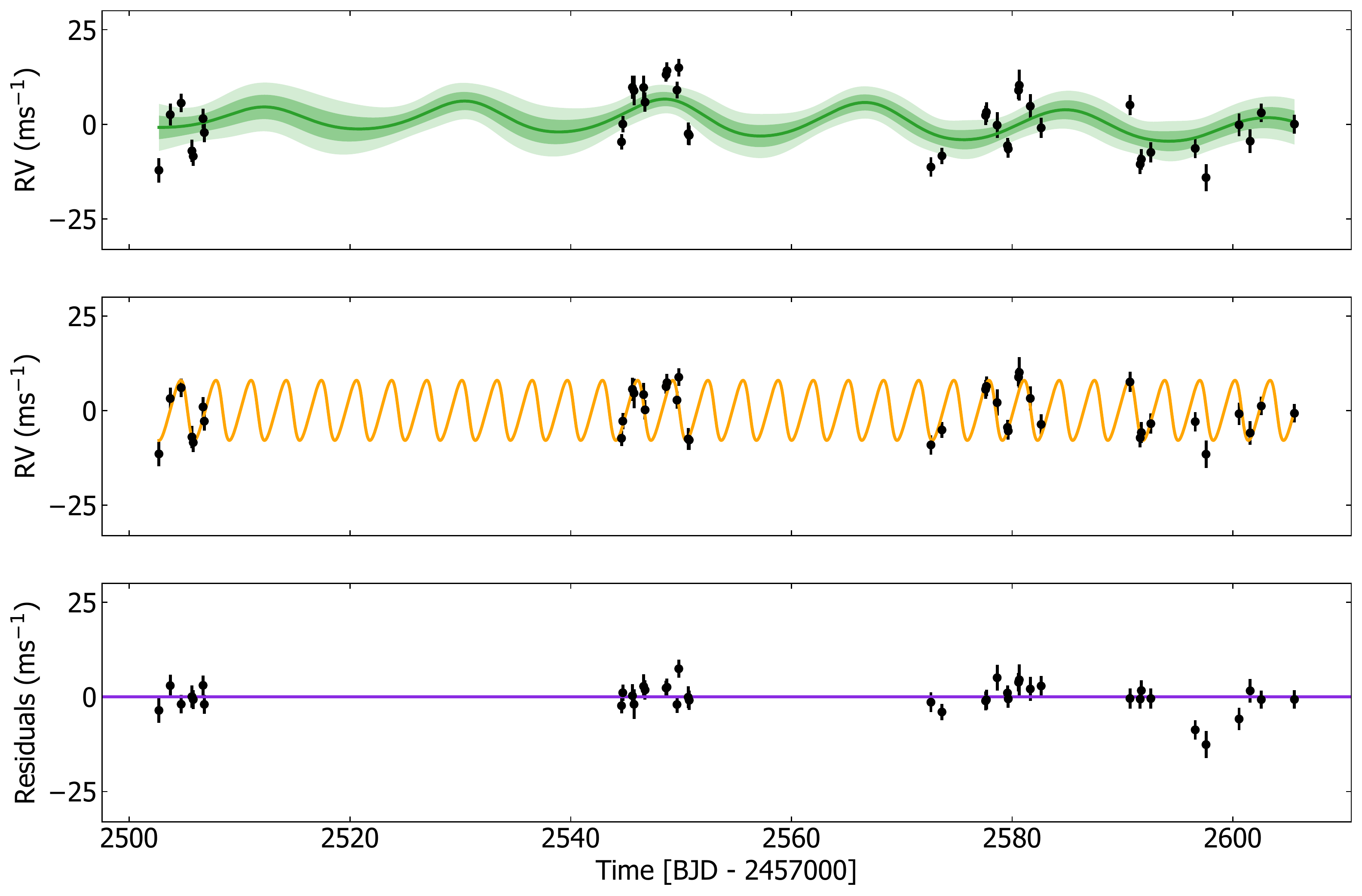}
    \caption{\textbf{Top panel:} \harps\ radial velocity data for \Tstar\ (black circles) with formal uncertainties, and the GP model plotted in green with 1 and 2 standard deviations from the model either side. \textbf{Middle panel:} Radial velocity model of \Tplanetb\ (orange) with the GP model subtracted, overplotted with the \harps\ datapoints. \textbf{Bottom panel:} Residuals for the \harps\ data.}
    \label{fig:harpsfull}
\end{figure*}

\begin{figure}
    \centering
    \includegraphics[width=0.5\textwidth]{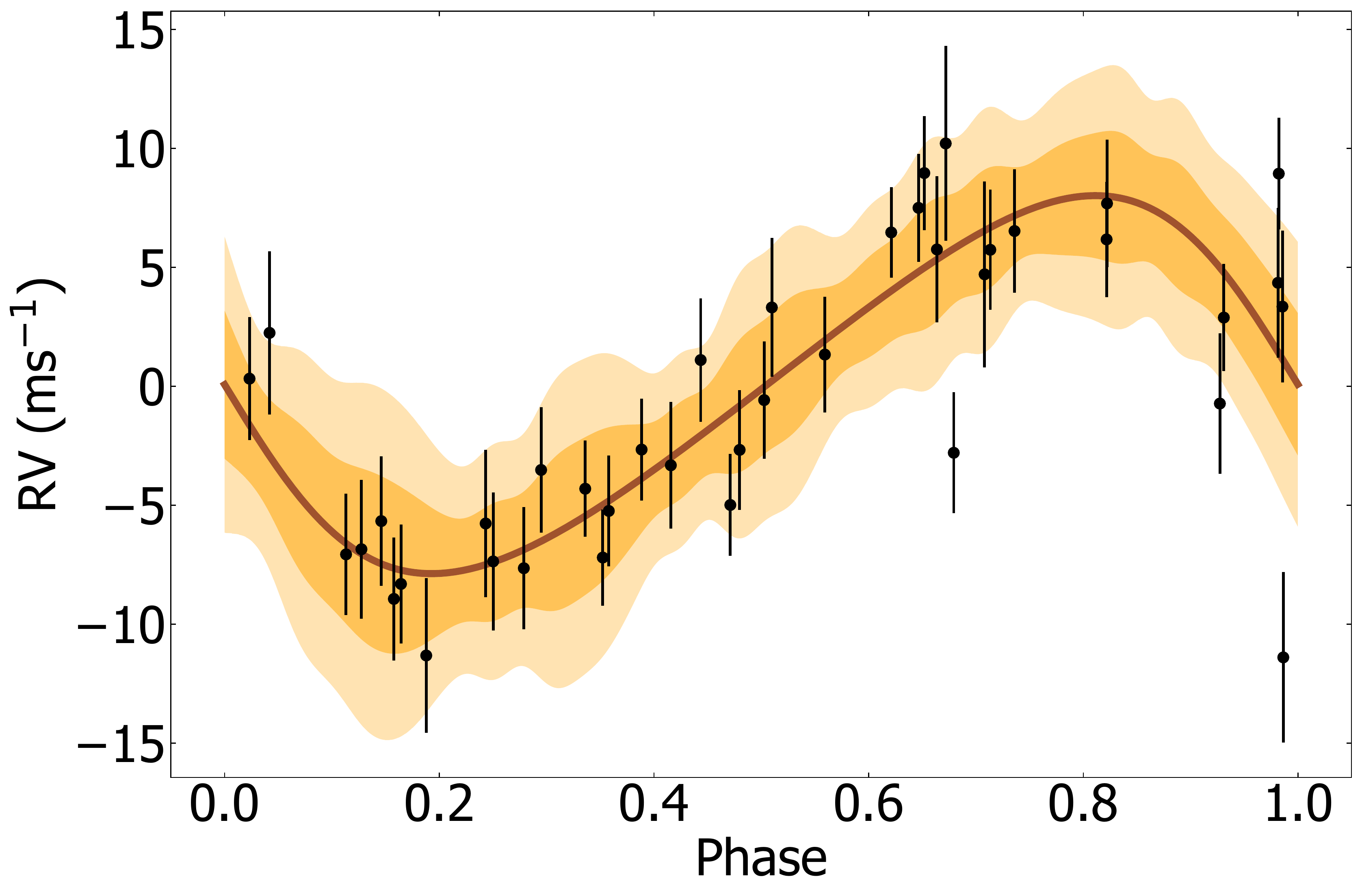}
    \caption{\harps\ radial velocity data for \Tstar, phase folded to a period corresponding to that of \Tplanetb\ and overplotted with the model in orange with 1 and 2 standard deviations from the model either side.}
    \label{fig:harpsphase}
\end{figure}

\begin{figure}
    \centering
    \includegraphics[width=0.5\textwidth]{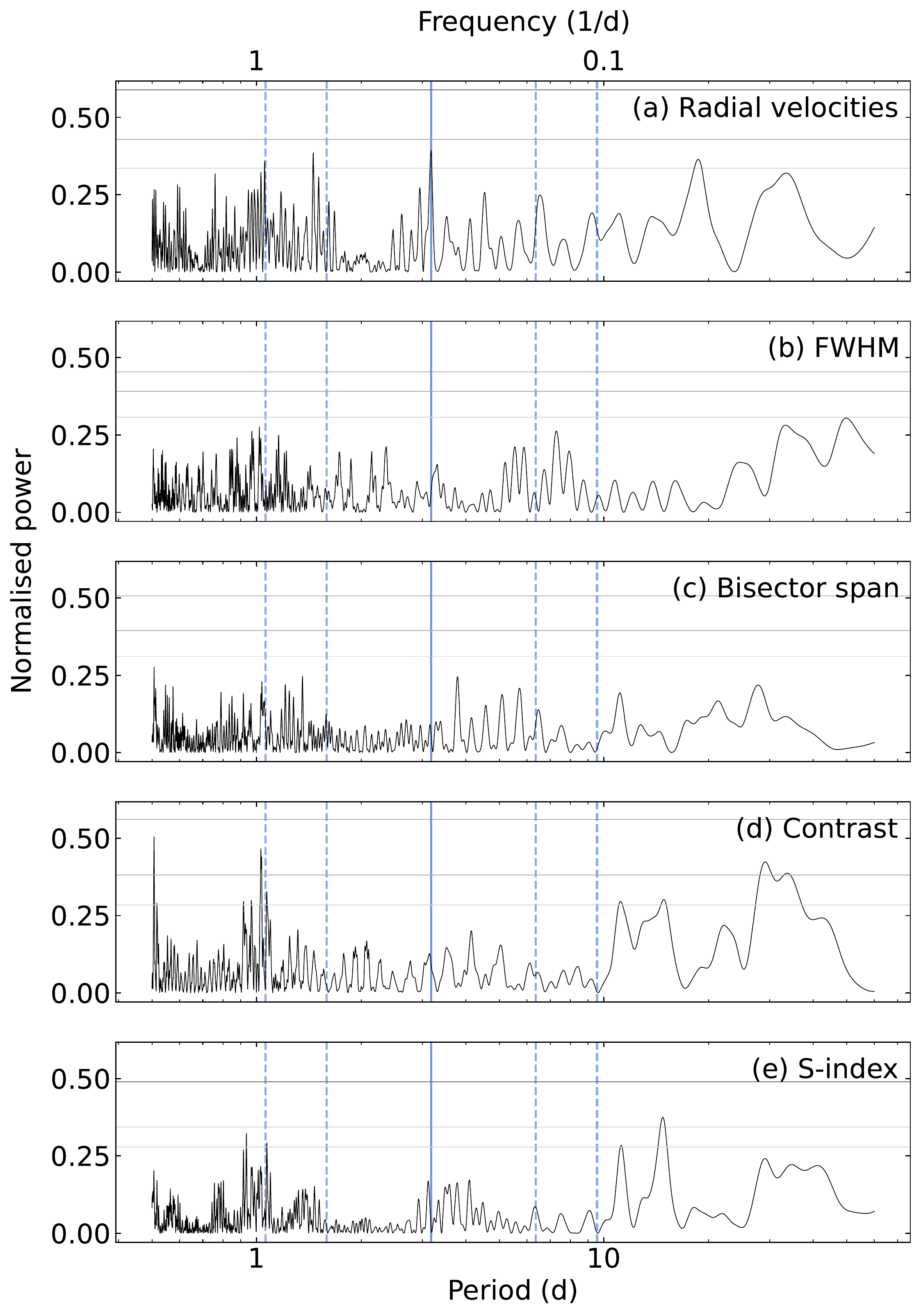}
    \caption{Periodograms for the \HARPS\ radial velocity data of the values presented in Table~\ref{tab:harpsobs}. The orbital period of \Tplanetb\ is marked with a solid vertical blue line, alongside two period aliases either side of this value marked as dashed vertical blue lines. The False Alarm Probabilities (FAP) are marked as the horizontal grey lines at 0.1, 1 and 10\% from top to bottom. \textbf{Panel 1}: Periodogram for the raw radial velocity data, with a peak at the orbital period value of \Tplanetb\ above the 0.1\% FAP line. \textbf{Panels 2-5}: Periodograms for the Full-Width Half Maximum (FWHM), bisector span, contrast and S-index\textsubscript{MW}.}
    \label{fig:rvperiods}
\end{figure}

It should also be noted that these data were obtained at relatively higher airmasses ($\sim$1.66) due to the on-sky position and the low declination of the target. These data are used in the RV (Radial Velocity) component of our joint model (Section~\ref{sec:jointmodel}), and the data are presented in Table~\ref{tab:harpsobs}. Our radial velocity data from \harps\ is shown in its entirety in Figure~\ref{fig:harpsfull} along with the fitted GP model and residuals, and the same data phase folded to the period of \Tplanetb\ is displayed in Figure~\ref{fig:harpsphase}. We present the Lomb-Scargle periodograms of the data presented in Table~\ref{tab:harpsobs} in Figure~\ref{fig:rvperiods} and compare each to the orbital period and period aliases of \Tplanetb. The periodograms of the additional activity indicators mentioned previously and their values are presented in Figure~\ref{fig:harpsappendix} and Table~\ref{tab:harpsappendix} respectively. We find additional periodicities above the 1\%\ False Alarm Probability (FAP) line in the CCF contrast, \textit{S}-index, Na-index and Ca-index, however due to the large uncertainty in the fitted stellar rotational period we cannot attribute these periodicities to this feature, and we encourage further monitoring of the spectroscopic radial velocity of the system.


\subsection{SOAR-HRCam speckle imaging}
\label{sec:imaging}

\begin{figure}
\centering
\includegraphics[width=\columnwidth]{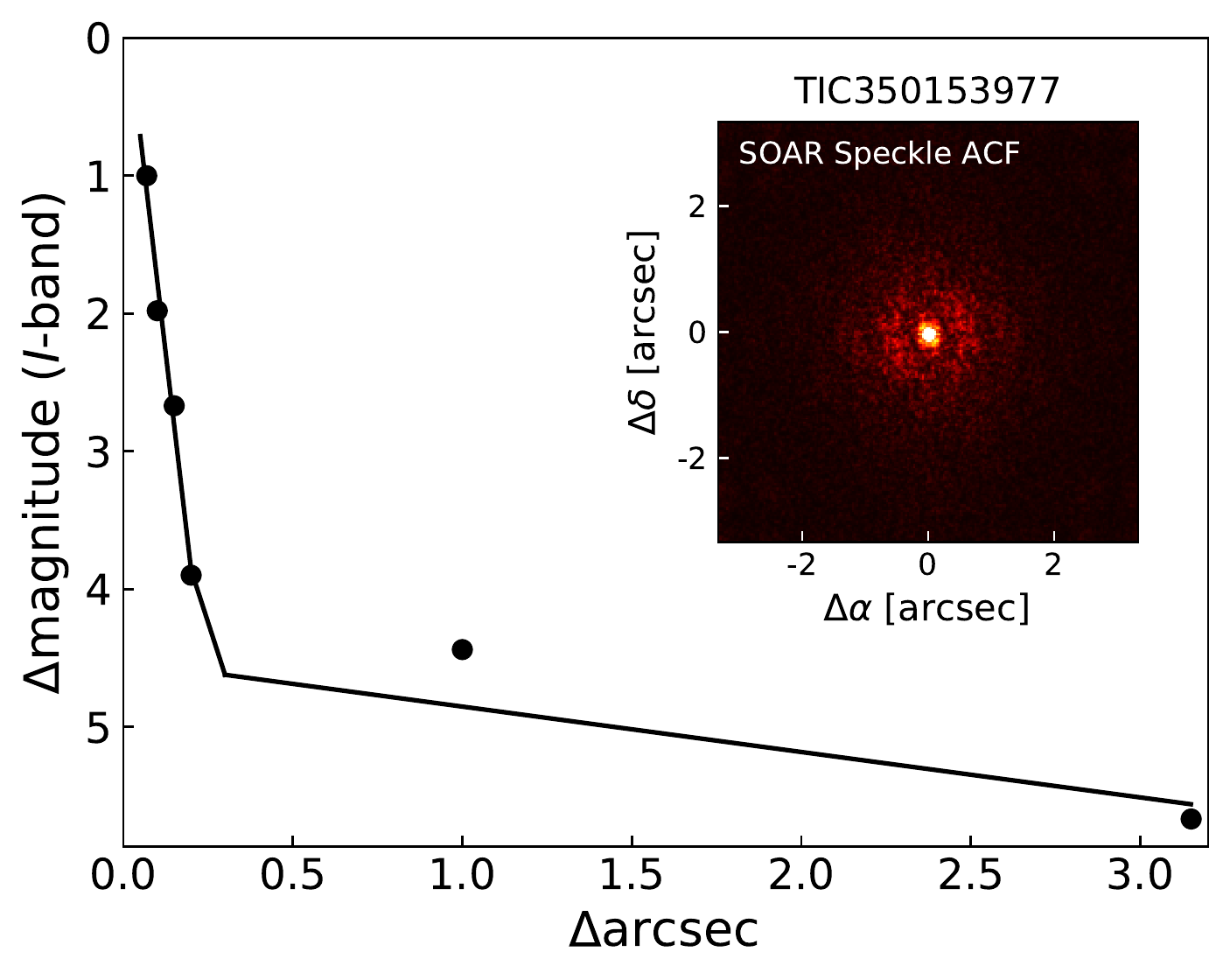}
\caption{\SOAR\ speckle imaging observations (insert), 5-$\sigma$ sensitivity limit and Auto-Correlation Functions (ACF) of \Tstar, showing no detection of close companions within 3\,\arcsec\ of the target.}
\label{fig:soar}
\end{figure}

To check for stellar companions to \Tstar\ that may be blended into the photometric data due to the comparatively large \tess\ pixel scale of 21\,\arcsec, \soar\ \citep[SOuthern Astrophysical Research telescope;][]{tokovinin2018} speckle imaging observations of the target were taken on 2019-10-16 in the Cousins-\textit{I} filter at a resolution of 36\,mas \citep{ziegler2020}. We show the 5-$\sigma$ sensitivity limit and Auto-Correlation Functions (ACF) for the observations in Figure~\ref{fig:soar}, and detect no contaminating sources within 3\,\arcsec\ of the target.


\section{Analysis} \label{sec:methods}

\subsection{Stellar analysis}
\label{sec:stellaranalysis}

To derive the stellar spectroscopic parameters  (\teff, $\log g$, microturbulence, [Fe/H]) we used \texttt{ARES+MOOG} following the same methodology described in \citet[][]{Sousa-21, Sousa-14, Santos-13}. The latest version of \texttt{ARES}\footnote{The latest version, \texttt{ARES v2}, can be downloaded at \url{https://github.com/sousasag/ARES}.} \citep{Sousa-07, Sousa-15} was used to measure the equivalent widths (EW) of iron lines on the combined spectrum of TOI-908. We then use a minimization process to find ionization and excitation equilibrium and converge to the best set of spectroscopic parameters. This process makes use of a grid of Kurucz model atmospheres \citep{Kurucz-93} and the radiative transfer code \texttt{MOOG} \citep{Sneden-73}. We also derived a more accurate trigonometric surface gravity using recent \gaia\ data following the same procedure as described in \citet[][]{Sousa-21}.

Stellar abundances of the elements were derived using the classical curve-of-growth analysis method assuming local thermodynamic equilibrium and with the same codes and models that were used for the stellar parameters determinations. For the derivation of chemical abundances of refractory elements we closely followed the methods described in \citep[e.g.][]{Adibekyan-12, Adibekyan-15, Delgado-17}. Abundances of the volatile elements, C and O, were derived following the method of \cite{Delgado-21, Bertrandelis-15}. All the abundance ratios [X/H] are obtained by doing a differential analysis with respect to a high S/N solar (Vesta) spectrum from \HARPS. The final abundances, shown in Table~\ref{tab:star_props_results}, are typical of a galactic thin-disk star. Moreover, we used the chemical abundances of some elements to derive ages through the so-called chemical clocks (i.e. certain chemical abundance ratios which have a strong correlation for age). We applied the 3D formulas described in table 10 of \citet{Delgado-19}, which also consider the variation in age produced by the effective temperature and iron abundance. The chemical clocks [Y/Mg], [Y/Zn], [Y/Ti], [Y/Si], [Y/Al], [Sr/Ti], [Sr/Mg] and [Sr/Si] were used from which we obtain a weighted average age of 4.6\,$\pm$\,1.5 Gyr. 


\subsection{Joint modelling}
\label{sec:jointmodel}

We use the \texttt{exoplanet} \citep{exoplanetdanforeman, exoplanetnew} code framework to jointly model the photometric and spectroscopic data for \Tstar\ detailed in Section~\ref{sec:obs}. \texttt{exoplanet} incorporates the packages \texttt{starry} \citep{starry}, \texttt{PyMC3} \citep{exoplanet:pymc3} and \texttt{celerite} \citep{celerite}. Our photometry data subset contains all observations presented in Table~\ref{tab:photobs}, and our RV data subset consists of the \harps\ observations presented in Section~\ref{sec:harpsrv}, all of which have been converted to the TBJD time system (\tess\ Barycentric Julian Date; BJD - 2457000) for uniformity. To create a complete model which incorporates all observations and the GP (Gaussian Process) models for each, outlined further in Sections~\ref{sec:jointmodelphotometry} and \ref{sec:jointmodelrv} below, we obtain our initial fit values from the maximum log probability of the \texttt{PyMC3} model, and then use these values as the starting point to draw samples from the posterior distribution using a NUTS (No U-Turn Sampler) variant of the HMC \citep[Hamilton Monte Carlo;][]{mcmcnuts} algorithm. We use a burn-in of 4000 samples which are discarded, 4000 steps and 10 chains, which gives our model good convergence without excessive computation time. The prior distributions implemented and their resulting fit values can be found in Table~\ref{tab:priorsstar} for \Tstar, and Table~\ref{tab:priorsb} for \Tplanetb. The priors and resulting fit values for the \tess\ GP parameters can be found in Table~\ref{tab:priorsphot}.


\subsubsection{Light curve detrending}
\label{sec:jointmodelphotometry}

To remove residual effects of stellar variability on the \TESS\, light curves that were not fully removed by the PDCSAP algorithm, we use a GP (Gaussian Process) model on each of the six Sectors of data using the \texttt{celerite} and \texttt{PyMC3} packages. This GP model is defined by three hyper-parameters for each \tess\ Sector, with log(\textit{s}2) describing the excess white noise in the data, and log($\omega$0) and log(S$\omega$4) representing the non-periodic components of stellar variability in the light-curves \citep{exoplanet:pymc3}. These parameters are passed into the SHOTerm kernel in the \texttt{exoplanet} framework, which represents a stochastically-driven simple harmonic oscillator \cite{exoplanetdanforeman}.
The effect of these GP models can be seen in Figure~\ref{fig:gpmodels}, and the corresponding equation is presented in Equation~\ref{eq:tessgp}, where $S_{0}\omega_{0}^{4}$ represents the \textit{S$\omega$}4 term as described above and $Q = \frac{1}{\sqrt{2}}$,

\begin{equation} \label{eq:tessgp}
    S(\omega) = \sqrt{\frac{2}{\pi}} \frac{S_{0}\omega_{0}^{4}}{(\omega ^{2}-\omega _{0}^{2})^{2} + \omega _{0}^{2}\omega ^{2}/Q}.
\end{equation}

We normalise each of our follow-up photometry observations from \lco\ by dividing each light-curve by the median of the out-of-transit flux and subtracting the mean of the out-of-transit flux. We find no need to apply a GP model to these light-curves due to the shorter baselines of the observations over the course of a single transit.

Each of the transits from the photometry data are modelled within the \texttt{exoplanet} code as a Keplerian orbit following the formalisation of \citet{exoplanet:kipping13}, defined by the stellar parameters of radius (\rstar) and mass (\mstar) in Solar units, and the planetary parameters of orbital period (\textit{P}) in days, central transit ephemeris (\tc) in TBJD (BJD-2457000), impact parameter (\textit{b}), eccentricity (\textit{e}) and argument of periastron (\textit{$\omega$}) in radians, including the limb-darkening coefficients $u_1$ and $u_2$. A set of transit models for each data set is generated using the \texttt{starry} package contained in \texttt{exoplanet}, which also incorporates the planetary radius (\rpl) and exposure times for each instrument (see Table~\ref{tab:photobs}).


\subsubsection{Radial velocity (RV) detrending}
\label{sec:jointmodelrv}

\begin{figure}
    \centering
    \includegraphics[width=0.5\textwidth]{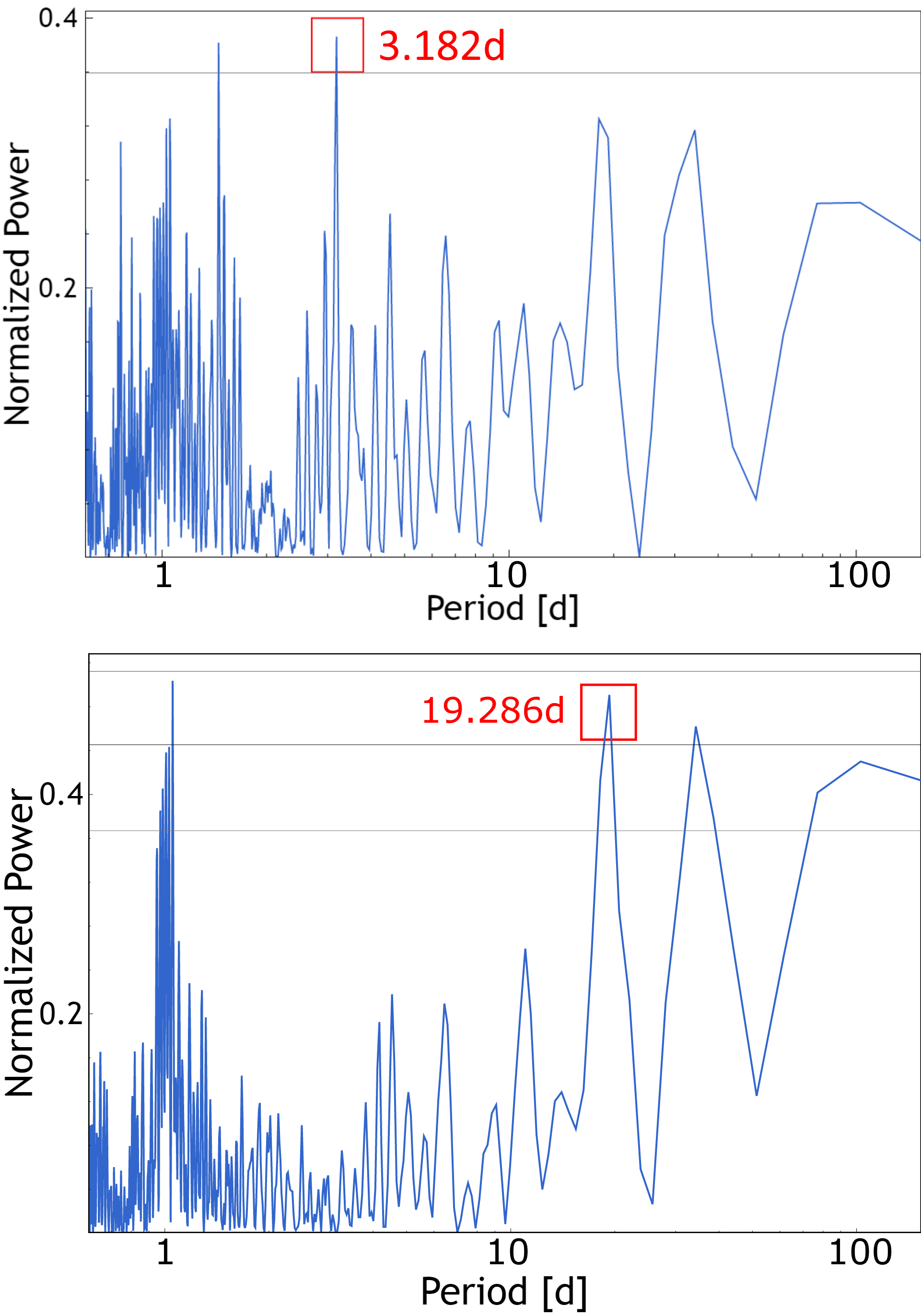}
    \caption{\textbf{Top panel:} Periodogram of the \harps\ radial velocity data for \Tstar\ showing the signal of \Tplanetb, marked with a red box. The horizontal line represents the 1\% FAP level. \textbf{Bottom panel:} Periodogram of the \harps\ radial velocity data for \Tstar\ with the signal of \Tplanetb\ removed, showing the remaining stellar rotational signal. The horizontal lines represent the 0.1\%, 1\% and 10\% FAP levels.}
    \label{fig:dace}
\end{figure}

We first examine the \harps\ radial velocity data (RMS 0.0048, mean error 2.79\,\ms) using the DACE\footnote{DACE is accessible at: \url{https://dace.unige.ch/}.} platform, in which we find two significant periodic signals above the level of the 1\% analytical FAP (False Alarm Probability) that are not aliases of each other (Figure \ref{fig:dace}). We find the planetary signal of \Tplanetb\ on a period of 3.182\,days with a predicted radial velocity semi-amplitude \textit{K} of 7.72\,\ms, and a secondary periodic signal at 19.29\,days with a semi-amplitude of 7.25\,\ms\ after removal of the planetary signal in DACE. We also calculate a rotational period of $20.53^{+6.77}_{-4.06}$\,days based on a measured \vsini\ of $2.56\,\pm\,0.64$\,\kms (Table~\ref{tab:star_props_results}, Section~\ref{sec:stellaranalysis}), and as the peak periodic signal from DACE falls within this tolerance, we attribute it to stellar rotational modulation (\textit{P}\textsubscript{rot}).

As with the photometric \tess\ data, we also apply a GP model to the \harps\ data to account firstly for the stellar rotational period, and secondly for any residual instrumental or noise effects that are not accounted for in the data reduction. The prior for \textit{P}\textsubscript{rot} is set up as a normal distribution around the predicted value of 20.53\,days with a standard deviation of 7\,days (Table~\ref{tab:priorsstar}), based on the analysis from DACE. Our quasi-periodic GP kernel used is identical to that of \citet{hawthorn2022} and \citet{ares}, which is a combination of the \texttt{Periodic} and \texttt{ExpQuad} (squared exponential) kernels available from the \texttt{PyMC3} package \citep{pymc3}, multiplied to create our final kernel in Equation~\ref{eq:rvkernel}. This final kernel is parameterised by the GP amplitude $\eta$, the stellar rotation period \textit{P}\textsubscript{rot}, the timescale of active region evolution \textit{l}\textsubscript{\textit{E}} and the smoothing parameter \textit{l}\textsubscript{\textit{P}}:

\begin{equation} \label{eq:rvkernel}
    k(x,x') = \eta^2 \exp \left(-\frac{\sin^2(\pi |x-x'| \frac{1}{P_{\rm rot}})}{2l_P^2} - \frac{(x-x')^2}{2l_E^2} \right).
\end{equation}

We first find predicted values for the radial velocity of \Tplanetb\ at each timestamp in the \harps\ data, which uses a uniform prior between 0 and 10~\ms\ for the semi-amplitude \textit{K} of the planet signal. We also fit for the instrumental offset of \harps\, to account for the differences in RV zero points between instruments, and any other residual effects not incorporated into the \harps\ formal uncertainties or the GP model. Our noise model adds jitter noise in quadrature with the nominal RV uncertainties in Equation~\ref{eq:rvnoise},

\begin{equation} \label{eq:rvnoise}
    \sigma^2 = {\sigma_{0}}^2 + {\sigma_{i}}^2 ,
\end{equation}

where $\sigma_{0}$ is the RMS of the jitter noise and $\sigma_{i}$ is the nominal uncertainty of the i-th radial velocity measurement. The priors and resulting fit values for the \harps\ radial velocity GP can be found in Table~\ref{tab:priorsrv}.


\section{Results and Discussion} \label{sec:results}
The results of our joint modelling show that \Tplanetb\ is a sub-Neptune with a mass of \TMassbshort\,\mearth\ and a radius of \TRadiusbshort\,\rearth\ orbiting a G-dwarf star in a close-in \Tperiodbshort~day orbit. Our final set of parameters for \Tplanetb\ can be found in Table~\ref{tab:planet_props}.

\begin{center}
\begin{table}
    \centering
    \caption{Stellar parameters of \TStar.}
    \label{tab:star_props_results}
    \begin{threeparttable}
    \begin{tabularx}{0.96\columnwidth}{ l  l X }
    \toprule
    \textbf{Property (unit)} & \textbf{Value} & \textbf{Source} \\
    \hline
    Mass (\msun)    & \Tstarmassexo & \texttt{exoplanet} \\
    Radius (\rsun)  & \Tstarradiusexo   & \texttt{exoplanet} \\
    Density (\gccc) & \Tstardensityexo  & \texttt{exoplanet} \\
    $P_{\rm rot}$ (days)   & \Tstarperiodexo   & \texttt{exoplanet} \\
    LD coefficient \textit{u\textsubscript{1}}  & \ldu  & \texttt{exoplanet} \\
    LD coefficient \textit{u\textsubscript{2}}  & \ldv  & \texttt{exoplanet} \\
    \logg\ &\Tloggporto  &\texttt{ARES + MOOG + \gaia} \\
    \teff\ (K)   &\Teffporto  &\texttt{ARES + MOOG} \\
    \vsini\ (\kms) &\Tstarvsiniporto &\texttt{ARES + MOOG}\\
    \vturb\ (\kms)  & \Tstarvturbporto &\texttt{ARES + MOOG}\\
    Age (Gyr)   &\Tstarageporto & Chemical clocks \\
    \hline
    \multicolumn{3}{l}{\textbf{Stellar abundances}} \\
    \feh\,(dex) &\Tstarfehporto &\texttt{ARES + MOOG} \\
    \ch\,(dex) &\Tstarchporto &\texttt{ARES + MOOG} \\
    \oh\,(dex) &\Tstarohporto &\texttt{ARES + MOOG} \\
    \nah\,(dex) &\Tstarnahporto &\texttt{ARES + MOOG} \\
    \mgh\,(dex) &\Tstarmghporto &\texttt{ARES + MOOG} \\
    \alh\,(dex) &\Tstaralhporto &\texttt{ARES + MOOG} \\
    \sih\,(dex) &\Tstarsihporto &\texttt{ARES + MOOG} \\
    \tih\,(dex) &\Tstartihporto &\texttt{ARES + MOOG} \\
    \nih\,(dex) &\Tstarnihporto &\texttt{ARES + MOOG} \\
    \cuh\,(dex) &\Tstarcuhporto &\texttt{ARES + MOOG} \\
    \znh\,(dex) &\Tstarznhporto &\texttt{ARES + MOOG} \\
    \srh\,(dex) &\Tstarsrhporto &\texttt{ARES + MOOG} \\
    \yh\,(dex) &\Tstaryhporto &\texttt{ARES + MOOG} \\
    \zrh\,(dex) &\Tstarzrhporto &\texttt{ARES + MOOG} \\
    \bah\,(dex) &\Tstarbahporto &\texttt{ARES + MOOG} \\
    \ceh\,(dex) &\Tstarcehporto &\texttt{ARES + MOOG} \\
    \ndh\,(dex) &\Tstarndhporto &\texttt{ARES + MOOG} \\

    \bottomrule
    \end{tabularx}
    \begin{tablenotes}
    \item Sources: \texttt{exoplanet} \citep{exoplanetdanforeman, exoplanetnew}, \texttt{ARES} \citep{Sousa-15}, \texttt{MOOG} \citep{Sneden:1973, Kurucz:1993}, \gaia\ \citep{GAIA_DR3}
    \end{tablenotes}
    \end{threeparttable}
\end{table}
\end{center}


\begin{table}
    \caption{Parameters of \Tplanetb.}
    \label{tab:planet_props}
\begin{tabular}{ll}
    \toprule
        \textbf{Property (unit)} & \textbf{Value} \\
        \hline
            Catalog identifier & TOI-908.01 \\
            Period (days) & \Tperiodb \\
            Mass (\mearth) & \TMassb \\
            Radius (\rearth) & \TRadiusb \\
            Density (\gccc) & \Tdensityb \\
            \rpl/\rstar & \Trorb \\
            \tc\ (TBJD) & \Tcb \\
            T1-T4 duration (hours) & \TTDurfullb \\
            T2-T3 duration (hours) & \TTDurcutb \\
            Impact parameter & \Timpactb \\
            \textit{K} (\ms) & \Tkb \\
            Inclination ($^{\circ}$) & \Tincb \\
            Semi-major axis (AU) & \Taub \\
            Temperature \teq\ (K) $^*$  & \Teqb \\
            Insolation flux (\fsun) & \Tfluxb \\
            Eccentricity & \Teccb \\
            Argument of periastron ($^{\circ}$) & \Tomegab  \\
        \bottomrule
    \end{tabular}
    \begin{tablenotes}
            \item $^*$ Assuming Albedo\,=\,0 and uniform surface temperature
    \end{tablenotes}
\end{table}






\subsection{Density}
\label{results:MR}

\begin{figure}
    \centering
    \includegraphics[width=0.5\textwidth]{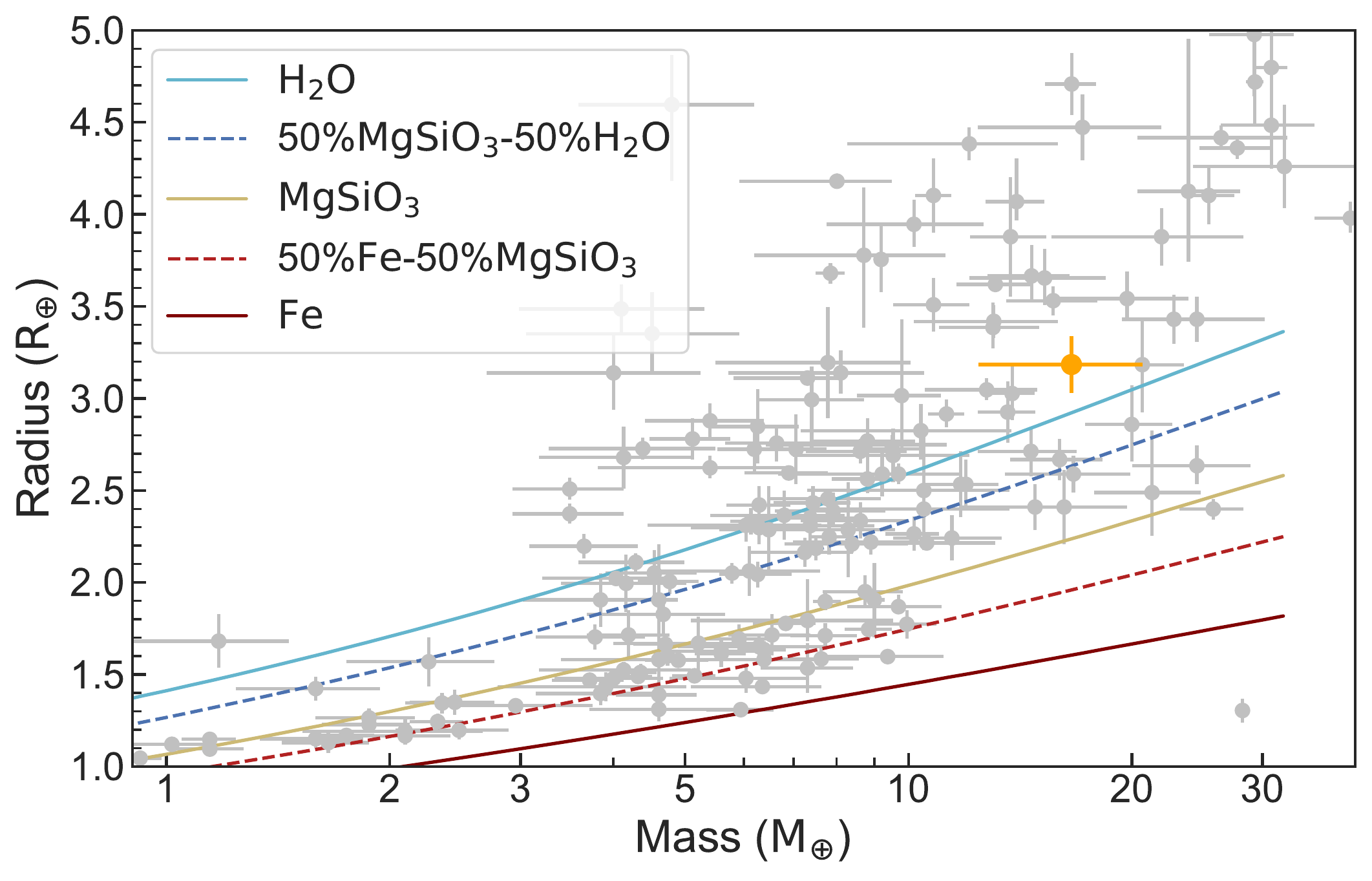}
    \caption{Mass-radius diagram showing the position of \Tplanetb\ (orange) amongst the population of exoplanets from the \textit{TEPCAT} catalog \citep{tepcat} (grey points), overplotted with composition models from \citet{zeng}.}
    \label{fig:massradius}
\end{figure}

In Figure~\ref{fig:massradius} we plot the position of \Tplanetb\ relative to the sample of planets from the \textit{TEPCAT} catalog \citep{tepcat}, and relative to composition models obtained from \citep{zeng} using the open-source code \texttt{fancy-massradius-plot}\footnote{\url{https://github.com/oscaribv/fancy-massradius-plot}}. \Tplanetb\ sits above the 100\%\ water planetary composition model amongst a more sparse population. The mass, radius and resulting bulk density of \Tplanetb\ (Table~\ref{tab:planet_props}) is similar to that of other known hot Neptune planets.

\subsection{Position of the planet in the Neptune desert}
\label{sec:neptunedesert}

\begin{figure}
    \centering
    \includegraphics[width=0.5\textwidth]{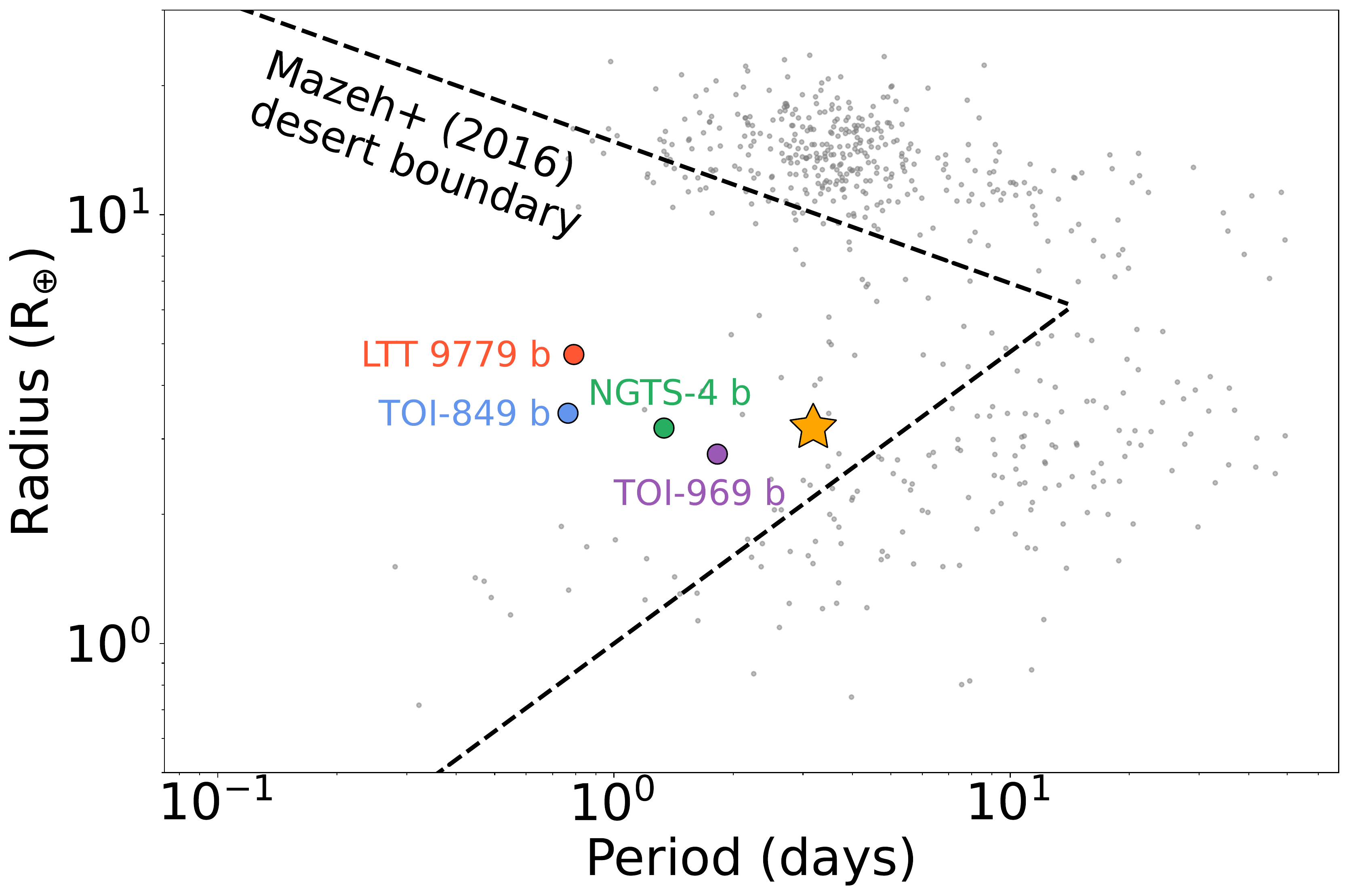}
    \caption{Plot showing the position of \Tplanetb\ (orange star) relative to the Neptune desert boundary of \citet{mazeh2016}, represented against the population of planets with measured periods and radii from the NASA Exoplanet Archive, and four confirmed planets in this parameter space: LTT 9779\,b \citep{jenkins2020} (red circle), TOI-849\,b \citep{Armstrong_2019} (blue circle), NGTS-4\,b \citep{west2019} (green circle) and TOI-969\,b \citep{lillobox2022} (purple circle).}
    \label{fig:neptunedesert}
\end{figure}

In Figure~\ref{fig:neptunedesert} we plot the position of \Tplanetb\ relative to the sample of planets with measured periods, radii and masses from the NASA Exoplanet Archive, and relative to the Neptune desert boundary of \citet{mazeh2016}. It can be seen that \Tplanetb\ lies just within the desert boundary.

\subsection{Additional planets and TTVs}
\label{sec:additionalplanets}

\begin{figure}
    \centering
    \includegraphics[width=0.5\textwidth]{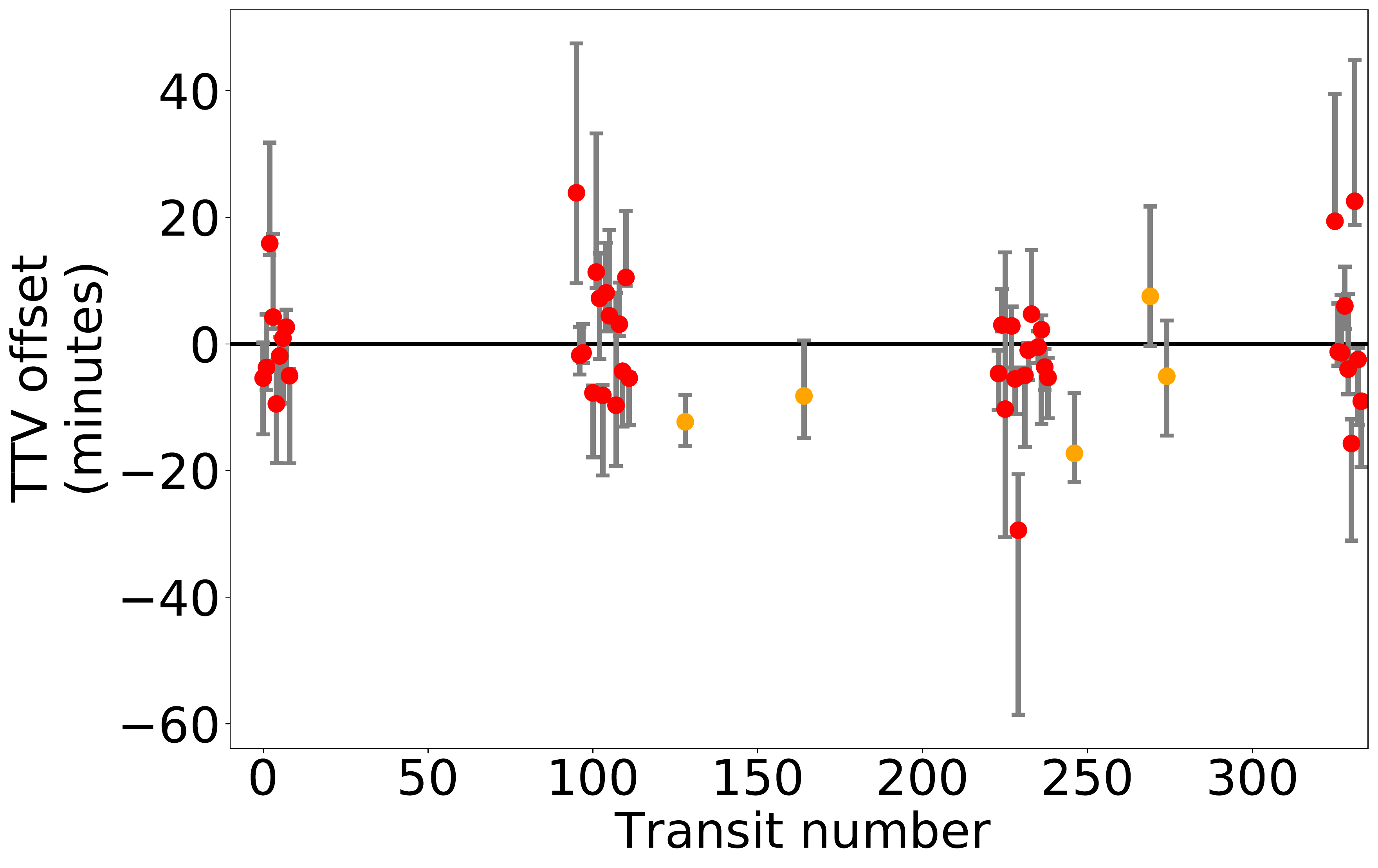}
    \caption{Transit Timing Varations (TTVs) for each transit of \Tplanetb\ from \lco\ (orange points), and each transit from \tess\ (red points), starting from transit number 0 taken as the first \tess\ transit in sector 1.}
    \label{fig:ttv}
\end{figure}

To evaluate the possible presence of additional planets in the system, both transiting and non-transiting, we implement additional searches of the \tess\ photometric data and the \harps\ spectroscopic radial velocity data. We perform a BLS \citep[Box Least Squares;][]{Kovacs2002} search on the \tess\ light curve data with the transit model for \Tplanetb\ subtracted, and find no evidence of additional significant signals above a False Alarm Probability (FAP) of 0.1 besides the predicted rotation period of 20.53~days. We search the \harps\ data using the DACE platform, which is capable of performing a periodogram search for additional periodic signals after the removal of the \Tplanetb\ radial velocity signal from the periodogram and the stellar rotation signal, and again we find no evidence for additional planets in the system. We also examine the individual transits from \tess\ and \lcogt\ with the \texttt{EXOFAST}\footnote{\url{https://exoplanetarchive.ipac.caltech.edu/cgi-bin/ExoFAST/nph-exofast}} platform to model the measured central transit times $T_{\rm cm}$ and compare them with the calculated $T_{\rm cc}$, and find that the measured transit times vary from the expected time by up to 29 minutes, consistent with our errors. We plot the TTVs for \Tplanetb\ in Figure~\ref{fig:ttv}.

\subsection{Internal structure modelling}
\label{sec:intstructure}

\begin{table}
    \caption{Internal structure of \Tplanetb\ from \protect\citet{rogers-owen-2021} models.}
    \label{tab:planet_structure}
\begin{tabular}{lll}
    \toprule
        \multicolumn{2}{c}{\textbf{Property (unit)}} \textbf{} & \textbf{Value}\\
        \hline
            Core radius & $R_{\rm core}$ (\rearth) & $2.31\pm0.17$ \\
            Core mass   & $M_{\rm core}$ (\mearth) & $16.0\pm4.0$ \\
            Envelope radius & $R_{\rm env}$ (\rearth) & $0.87\pm0.23$ \\
            Envelope mass fraction & $M_{\rm env}$/$M_{\rm p}$ & $0.022\pm0.010$ \\
        \bottomrule
\end{tabular}
\end{table}

We first estimate the internal structure of the planet assuming that it is comprised of a rocky core (Earth-like bulk density) surrounded by a H/He-rich envelope, following \citet{rogers-owen-2021}. The internal structure can thus be described by four parameters: the core radius $R_{\rm core}$ and mass $M_{\rm core}$, the envelope radius $R_{\rm env}$, and the envelope mass fraction $f_{\rm env}$, which is defined in Equation~\ref{eqn:fenv_def}:

\begin{equation} \label{eqn:fenv_def}
    f_{\rm env} = \frac{M_{\rm env}}{M_{\rm p}} = \frac{M_{\rm p} - M_{\rm core}}{M_{\rm p}},
\end{equation}

where $M_{\rm env}$ and $M_{\rm p}$ are the planet's envelope mass and total mass, respectively.
We make use of the empirical mass--radius relations for rocky planets by \citet{otegi-2020} to estimate the properties of the rocky core and relate core mass to core radius.
For the H/He envelope, we adopt the envelope structure model by \citet{chen-rogers-2016}, who provide a polynomial fit to their \texttt{MESA} simulations of the atmospheres of sub-Neptunes, in order to link envelope mass fraction and envelope radius.
These formulations together reduce the number of unknowns to two: the envelope mass fraction and the core mass,
which can be related to each other with the definition of envelope mass fraction above.
The resulting internal structure, shown in Table \ref{tab:planet_structure}, indicates an envelope mass fraction of $f_{\rm env} = 2.2\pm1.0\%$, typical of sub-Neptunes above the radius valley \citep{rogers-owen-2021}.

\begin{figure}
    \centering
    \includegraphics[width=0.45\textwidth]{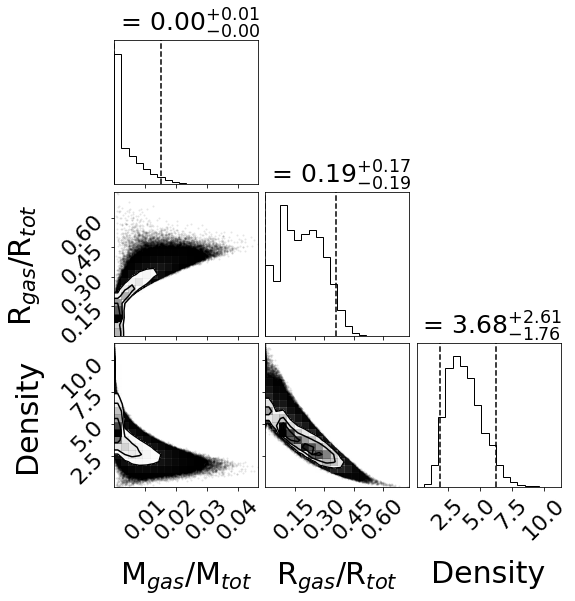}
    \caption{Corner plot of the derived internal structure parameters of \Tplanetb\ for the scenario with a water mass fraction of up to 0.5.}
    \label{fig:water}
\end{figure}

\begin{figure}
    \centering
    \includegraphics[width=0.45\textwidth]{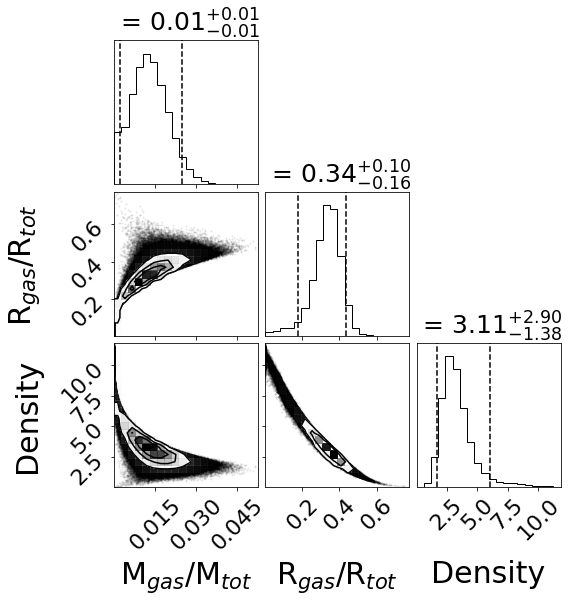}
    \caption{Corner plot of the derived internal structure parameters of \Tplanetb\ for the scenario with no water.}
    \label{fig:nowater}
\end{figure}

We perform an additional study of the internal structure of the planet with a Bayesian analysis, using the derived stellar and planetary properties. This method is described in details in \cite{Leleu20212} and has been applied on several systems such as L98-59 \citep{Demangeon2021}, TOI-178 \citep{Leleu20212} or Nu2 Lupi \citep{Delrez20212}. In the model of planetary interior, four layers are assumed: a inner core made of iron and sulfur, a mantle of silicate (Si, Mg and Fe), a water layer and a gaseous envelope of pure H-He. The core, mantle and water layers form the `solid' part of the planet and the thickness of the gaseous envelope depends on its mass and radius as well as the stellar age and irradiation \citep{Lopez2014}. The resulting planetary parameters are the mass fraction of each layer, the iron molar fraction in the core, the silicon and magnesium molar fraction in the mantle, the equilibrium temperature, and the age of the planet (equal to the age of the star). Uniform priors are used for these parameters, except for the mass of the gas layer which is assumed to follow a uniform-in-log prior, with the water mass fraction having an upper boundary of 0.5 \citep{Thiabaud2014,Marboeuf2014}. For more details related to the connection between observed data and derived parameters, we refer to \cite{Leleu2021}. Two scenarios have been investigated for \Tplanetb: the case with water (water mass fraction up to 0.5) and the case without water. The figures \ref{fig:water} and \ref{fig:nowater} refer to these two cases respectively. Both cases give results consistent with the planetary observables. The case with water leads to an atmosphere representing 18\% of the total radius, whereas the case without water produces a thicker gaseous envelope of almost 35\% of the total radius.

\begin{table}
    \caption{Internal structure of \Tplanetb\ from \protect\citet{Leleu2021} models.}
    \label{tab:planet_structure2}
\begin{tabular}{llll}
    \toprule
        \multicolumn{2}{c}{\textbf{Property (unit)}} \textbf{} & \textbf{Water} & \textbf{No water} \\
        \hline
            Total radius & $R_{\rm tot}$ (\rearth) &  $2.9\pm0.47$ & $3.04\pm0.46$\\
            Total mass   & $M_{\rm tot}$ (\mearth) &  $16.5\pm4.45$ & $16.05\pm4.79$\\
            Envelope radius & $R_{\rm env}$ (\rearth) &  $0.54\pm0.6$ & $1.02\pm0.52$\\
            Envelope mass fraction & $M_{\rm env}$/$M_{\rm p}$ &  $0.002\pm0.010$ & $0.010\pm0.010$\\
        \bottomrule
\end{tabular}
\end{table}


\subsection{Evaporation history of the planet}
\label{sec:evolution}

The Neptune desert and the radius valley are consistent with being the result of the evaporation of the atmospheres of sub-Neptunes \citep{lopez-fortney-2013, lopezfortney2014, Owen2013, jin-2014}. The underlying cause of evaporation is still under debate, though, and several mechanisms have been proposed. One such mechanism is photoevaporation, in which X-ray and extreme ultraviolet radiation (together, XUV), originating from the host star, provides the energy for evaporation \citep{watson-1981, erkaev-2007}. Photoevaporation has been shown to reproduce both the radius valley \citep{rogers-owen-2021} as well as the lower boundary of the Neptune desert \citep{owen-lai-2018}.
\citet{Ginzburg2016} proposed an alternative mechanism for evaporation, core-powered mass loss, where the energy for atmospheric escape is provided by the internal thermal energy of the planet, and \citet{Gupta19:core-powered} showed that it can also reproduce the radius-period valley.
The existence of the radius valley has also been attributed to formation mechanisms \citep{Zeng19:water-valley} and impacts by planetesimals \citep{Wyatt20:valley-from-impacts}.

To study the evaporation history of the planet under this hypothesis, three ingredients are necessary: (1) a description of the XUV history of the star, (2) an envelope structure model \citep[here we adopt ][]{chen-rogers-2016}, and (3) a mass loss model that relates the incident X-ray flux on the planet to the amount of mass lost.

\begin{figure*}
    \centering
    \includegraphics[width=\textwidth]{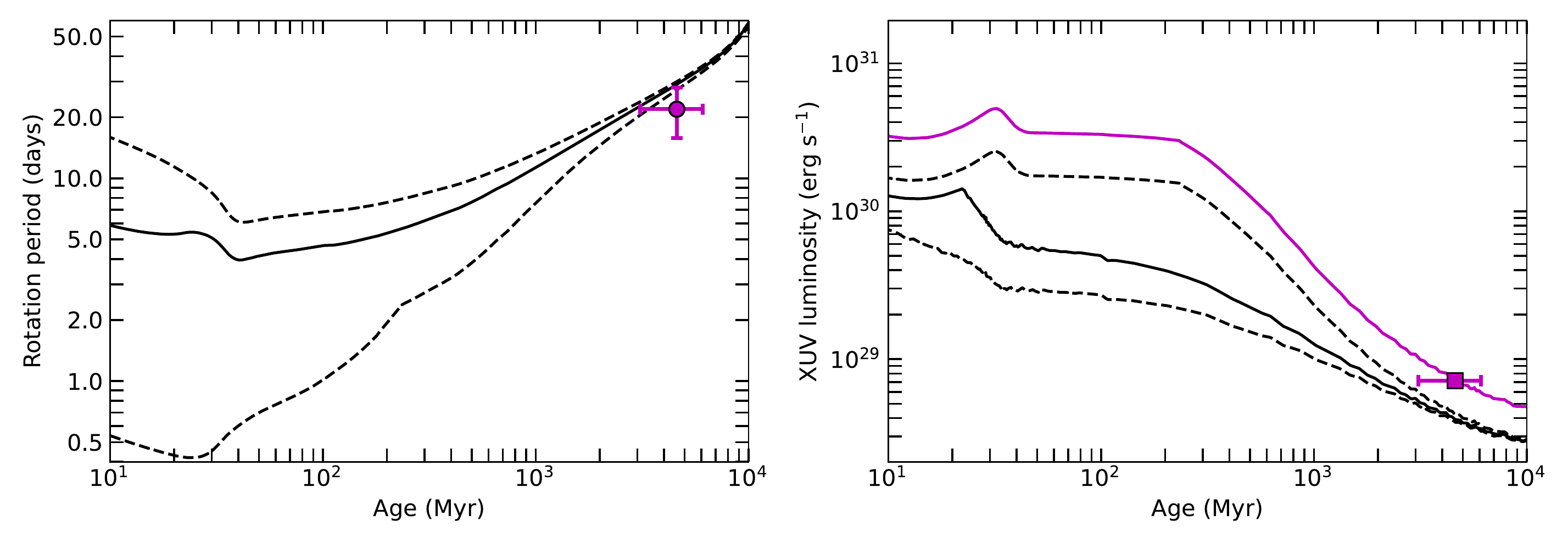}
    \caption{\textbf{Left panel:} Plot showing the evolution of the rotation period of a $0.95\,\text{M}_\odot$ star using the initial rotation periods of the population mean (solid line) and the $2\sigma$ spread (dashed lines), following the model by \citet{johnstone-2021}. \TStar, with a rotation period faster than expected for its age, is plotted as a purple circle. \textbf{Right panel:} XUV evolution tracks of \TStar\ based on the rotational histories shown on the left panel. together with the expected XUV luminosity of \TStar, based on its rotation period (purple square), and the fitted XUV track to that value (purple line).}
    \label{fig:star-evo}
\end{figure*}

The XUV history of a star can be estimated from its rotational history, as the two quantities are linked via the rotation--activity relation, where faster rotators produce higher X-ray fluxes \citep{wright-2011,wright-2018}. The X-ray luminosity of a star also declines with age, as stars spin down due to angular momentum loss through stellar winds \citep{jackson-2012, tu-2015, johnstone-2021}. We adopt the rotational evolution models by \citet{johnstone-2021}, who model the rotational spread and evolution of FGKM stars as a function of age and stellar mass. Figure \ref{fig:star-evo} shows the population mean and $2\sigma$ rotation tracks of a star of the mass of \TStar, and their corresponding XUV tracks. Furthermore, we adopt the mass loss formulation by \citet{kubyshkina-fossati-2021}, based on hydrodynamic simulations of planetary atmospheres in the context of photoevaporation.

The model by \citet{johnstone-2021} predicts that a 0.95~\msun\ star has an expected rotation period of 29~days, with a $2\sigma$ spread between 27 and 30~days, at an age of 4--5~Gyr.
As shown in Figure \ref{fig:star-evo} (left hand panel), \TStar,with an age of $4.6\pm1.5$\,Gyr, has a rotation period of $21.932\pm6.167$\,days, which
is a factor of $\sim$\,1.5 times faster than predicted by the model, making the star an unusually fast rotator for its age.
The $1\sigma$ uncertainties in the age and spin period, however,
allow for younger and slower scenarios consistent with the spin evolution models.
To model the XUV history of the star, we estimate the current X-ray luminosity using the rotation-activity relation, and use it to scale the model's $2\sigma$ upper limit X-ray track to fit our estimate, as shown in Figure \ref{fig:star-evo} (right hand panel). Furthermore, we estimate the extreme ultra-violet (EUV) luminosity of the star using the empirical relations by \citet{king-2018}.

\begin{figure*}
    \centering
    \includegraphics[width=\textwidth]{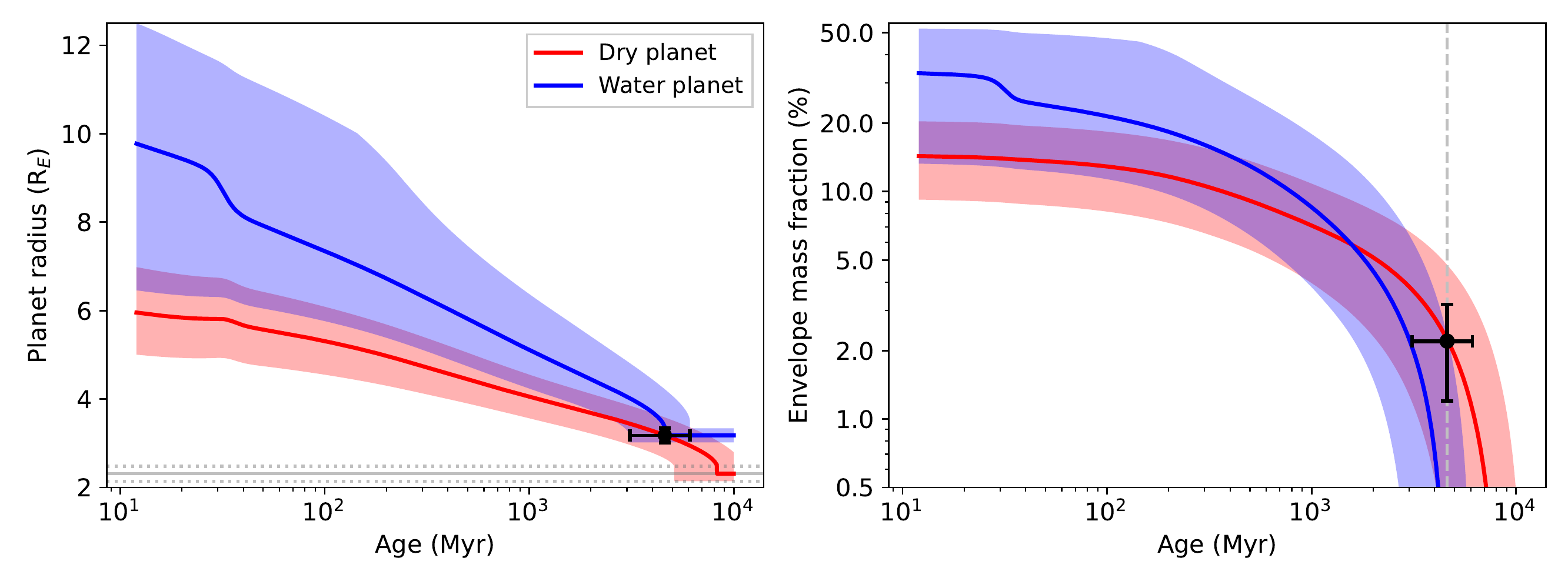}
    \caption{
    \textbf{Left panel}: plot showing the evolution of the radius of \Tplanetb\ using two models for the internal structure, one with a rocky core surrounded by gaseous atmosphere (red), and another with water instead of gas (blue). The shaded regions represent the spread in evaporation histories allowed by the $1\sigma$ uncertainties on the planet's mass, radius, and age. The present-day age and radius are shown with a black marker, and the rocky core radius (for the dry planet scenario) is shown as a horizontal grey line.
    \textbf{Right panel}: plot showing the evolution of the envelope mass fraction of \Tplanetb, akin to the left panel. The present-day envelope mass fraction for the dry planet scenario is shown as a black circle, and the current age of the planet is plotted as a grey vertical line.}
    \label{fig:planet-evo}
\end{figure*}

We simulate the evaporation history of \Tplanetb\ using the \texttt{photoevolver} code\footnote{\url{https://github.com/jorgefz/photoevolver}} \citep{fernandezfernandez2023}, which evolves the planet's envelope by following these steps iteratively: first the current XUV flux on the planet is drawn from the stellar evolution tracks (see Fig.\,\ref{fig:star-evo}), then the mass loss rate is obtained using the model by \citet{kubyshkina-fossati-2021}, which is then subtracted from the envelope mass, and finally the envelope size is recalculated using the \citet{chen-rogers-2016} model and the simulation jumps to the next timestep.
We run the simulation back to the age of 10 Myr and forward to 10 Gyr from the current age of $4.6\pm1.5$\,Gyr under the XUV irradiation history motivated by the star's measured spin period.
We simulated the evaporation past of the planet using the two internal structures we derived in Section~\ref{sec:intstructure}: a rocky core surrounded by a gaseous envelope consisting of $2.2\pm1.0$\% of its mass, which we designated the \textit{dry planet} scenario
(shown in Table~\ref{tab:planet_structure})
, and the Bayesian model by  \citet{Leleu2021} which accounts for the presence of water, and we designated the \textit{water planet} scenario (shown in Table~\ref{tab:planet_structure2}, left column).
The results, shown in Figure \ref{fig:planet-evo}, indicate that in the dry planet scenario, \Tplanetb\ could have started out
as a super-Neptune of radius 5--7~\rearth\ and envelope mass fraction 10--20\,\%.
On the other hand, we find that the evaporation past of the water scenario is fairly unconstrained, with initial states ranging from a 6.5\,\rearth\ planet akin to the initial state of the dry planet, to a puffy Jupiter-sized planet at 12\,\rearth\ with a significant envelope of mass fraction 50\%.
Furthermore, we also find that the planet's envelope
could be completely stripped by the age of 5--10\,Gyr in the case of the dry planet, whereas the tenuous gas layer on the water planet is removed in tens to a hundred Myr.

\section{Conclusion} \label{sec:conc}
We have presented the discovery of \Tplanetb, a hot sub-Neptune orbiting the G-type star \Tstar. We use a combination of transit photometry data from 6 sectors of \tess\ mission data at both 30\,minute and 10\,minute cadence, further follow-up observations from the \lcogt\ telescope network, radial velocity data from \harps, and \soar\ speckle imaging to rule out the presence of blended companions. We jointly model our transit photometry and radial velocity data, and find that the planet orbits a G-type star with a radius of \Tstarmassexoshort\,\msun\ and a radius of \Tstarradiusexoshort\,\rsun.

\Tplanetb\ has a radius of \TRadiusbshort\,\rearth\ and a mass of \TMassbshort\,\mearth, with an orbital period of \Tperiodbshort\,days. It lies within the parameter space of the period-radius diagram known as the `Neptune desert', an area of parameter space where a dearth of planets is seen in current demographics. The mean density \Tdensitybshort\,\gccc\ of \Tplanetb\ indicates an internal structure consisting of a core of radius $\sim$2.3~\rearth\ and mass of 16.0~\mearth, surrounded by an envelope of radius 0.87~\rearth\ with a mass fraction of 2.2\%. Our analysis of the evaporation history of the planet indicates that its host star is rotating faster than expected, and predict that the planet was previously a super-Neptune of radius 5--7~\rearth\ with an original envelope mass fraction of 10--20\%. Our models show the envelope will continue to evaporate in the future, and may be lost completely within the lifetime of the system. This planet is amenable to further follow-up with \jwst, as despite its relatively low Transmission Spectroscopy Metric \citep[TSM;][]{kempton2018} of \TSMb\ it is a fairly bright ($T$\,=\,\TTESSmagshort~mag) target undergoing atmospheric evaporation that lies close to the Continuous Viewing Zone of both \tess\ and \jwst\ for ease of observation scheduling. This target should also be considered for further spectroscopic evaporation measurements via the metastable He-line \citep{heline} with ground-based instruments, such as \textit{CARMENES} or \textit{NIRSPEC}.


\section*{Acknowledgements}

This work makes use of \texttt{tpfplotter} by J. Lillo-Box (publicly available in www.github.com/jlillo/tpfplotter), which also made use of the python packages \texttt{astropy}, \texttt{lightkurve}, \texttt{matplotlib} and \texttt{numpy}.

This research makes use of \texttt{exoplanet} \citep{exoplanet:exoplanet} and its dependencies \citep{exoplanet:agol20, exoplanet:arviz, exoplanet:astropy13,
exoplanet:astropy18, exoplanet:exoplanet, exoplanet:kipping13,
exoplanet:luger18, exoplanet:pymc3, exoplanet:theano}.

This work makes use of data from the European Space Agency (ESA) mission {\it Gaia} (\url{https://www.cosmos.esa.int/gaia}), processed by the {\it Gaia} Data Processing and Analysis Consortium (DPAC, \url{https://www.cosmos.esa.int/web/gaia/dpac/consortium}). Funding for the DPAC has been provided by national institutions, in particular the institutions participating in the {\it Gaia} Multilateral Agreement.

This paper includes data collected by the \TESS\ mission. Funding for the \TESS\ mission is provided by the NASA Explorer Program. Resources supporting this work were provided by the NASA High-End Computing (HEC) Program through the NASA Advanced Supercomputing (NAS) Division at Ames Research Center for the production of the SPOC data products. The TESS team shall assure that the masses of fifty (50) planets with radii less than 4 REarth are determined.

We acknowledge the use of public \TESS\ Alert data from pipelines at the \TESS\ Science Office and at the \TESS\ Science Processing Operations Center.

This research makes use of the Exoplanet Follow-up Observation Program website, which is operated by the California Institute of Technology, under contract with the National Aeronautics and Space Administration under the Exoplanet Exploration Program.

This paper includes data collected by the TESS mission that are publicly available from the Mikulski Archive for Space Telescopes (MAST).

This work makes use of observations from the LCOGT network. Part of the LCOGT telescope time was granted by NOIRLab through the Mid-Scale Innovations Program (MSIP). MSIP is funded by NSF.

This study is based on observations collected at the European Southern Observatory under ESO programme 1108.C-0697 (PI: Armstrong).

Based in part on observations obtained at the Southern Astrophysical Research (\soar) telescope, which is a joint project of the Minist\'{e}rio da Ci\^{e}ncia, Tecnologia e Inova\c{c}\~{o}es (MCTI/LNA) do Brasil, the US National Science Foundation’s NOIRLab, the University of North Carolina at Chapel Hill (UNC), and Michigan State University (MSU).

FH is supported by an STFC studentship.
JFF is supported by an STFC studentship.
AO is supported by an STFC studentship.
This work was supported by the UKRI Frontier research scheme (EP/X027562/1). DJA is supported by UKRI through the STFC (ST/R00384X/1) and EPSRC (EP/X027562/1).
This work was supported by Funda\c{c}\~{a}o para a Ci\^encia e a Tecnologia (FCT) and Fundo Europeu de Desenvolvimento Regional (FEDER) via COMPETE2020 through the research grants UIDB/04434/2020, UIDP/04434/2020.
JL-B acknowledges financial support received from "la Caixa" Foundation (ID 100010434) and from the European Unions Horizon 2020 research and innovation programme under the Marie Slodowska-Curie grant agreement No 847648, with fellowship code LCF/BQ/PI20/11760023. This research has also been partly funded by the Spanish State Research Agency (AEI) Project No.PID2019-107061GB-C61.
YA and JD acknowledge the support of the Swiss National Fund under grant 200020-172746.
This work has been carried out within the framework of the NCCR PlanetS supported by the Swiss National Science Foundation.
This publication was supported by an LSSTC Catalyst Fellowship to TD, funded through Grant 62192 from the John Templeton Foundation to LSST Corporation. The opinions expressed in this publication are those of the authors and do not necessarily reflect the views of LSSTC or the John Templeton Foundation.
MB acknowledges support from the STFC research grant ST/T000406/1.
ODSD is supported in the form of work contract (DL 57/2016/CP1364/CT0004) funded by FCT.
SH acknowledges CNES funding through the grant 837319.
HPO’s contribution has been carried out within the framework of the National Centre of Competence in Research PlanetS supported by the Swiss National Science Foundation under grants 51NF40\_182901 and 51NF40\_205606. The authors acknowledge the financial support of the SNSF.
This work was supported by FCT - Fundação para a Ciência - through national funds and by FEDER through COMPETE2020 - Programa Operacional Competitividade e Internacionalização by these grants: UID/FIS/04434/2019; UIDB/04434/2020; UIDP/04434/2020; PTDC/FIS-AST/32113/2017 \& POCI-01-0145-FEDER-032113; PTDC/FIS-AST/28953/2017 \& POCI-01-0145-FEDER-028953; PTDC/FIS-AST/28987/2017 \& POCI-01-0145-FEDER-028987; PTDC/FIS-AST/30389/2017 \& POCI-01-0145-FEDER-030389.
PW is supported by an STFC consolidated grant (ST/T000406/1).


\section*{Data Availability}
The \tess\ data is accessible via the MAST (Mikulski Archive for Space Telescopes) portal at \url{https://mast.stsci.edu/portal/Mashup/Clients/Mast/Portal.html}. Photometry and imaging data from \lco\ and \soar\ are accessible via the ExoFOP-\tess\ archive at \url{https://exofop.ipac.caltech.edu/tess/target.php?id=350153977}. The \texttt{exoplanet} modelling code and associated \texttt{python} scripts for parameter analysis and plotting are available upon reasonable request to the author.


\bibliographystyle{mnras}
\bibliography{bib.bib}


\appendix

\section{Priors and GP models} \label{sec:priors}


\begin{center}
\begin{table}
    \centering
    \caption{Global fit parameter prior function type and prior limits for \Tstar.}
    \label{tab:priorsstar}
    \begin{tabularx}{0.36\textwidth}{ l l l }
    \toprule
        \textbf{Parameter} &\textbf{Prior}  &\textbf{Value} \\
        \hline
        Baseline flux &$\mathcal{N}$(0, 1) & \\
        \mstar\ (\msun) &$\mathcal{N}$(0.95, 0.01, 0.95) & Table~\ref{tab:star_props_results}\\
        \rstar\ (\rsun) &$\mathcal{N}$(1.03, 0.03, 1.03) & Table~\ref{tab:star_props_results} \\
        \textit{P}\textsubscript{rot} (days) &$\mathcal{N}$(20.53, 7.00) & Table~\ref{tab:star_props_results} \\
        LD coefficient \textit{u\textsubscript{1}} &\citet{exoplanet:kipping13} & Table~\ref{tab:star_props_results} \\
        LD coefficient \textit{u\textsubscript{2}} &\citet{exoplanet:kipping13} & Table~\ref{tab:star_props_results} \\
        \bottomrule
    \end{tabularx}
    \begin{tablenotes}
    \item \textbf{Numbers in brackets represent:}
    \item (mean $\mu$, standard deviation $\sigma$, test value $\alpha$) for normal distribution $\mathcal{N}$($\mu$,$\sigma$,$\alpha$)
    \item Distributions for limb darkening coefficients \textit{u\textsubscript{1}} and \textit{u\textsubscript{2 }}are built into the \texttt{exoplanet} package and based on \citet{exoplanet:kipping13}
    \end{tablenotes}
\end{table}
\end{center}


\begin{center}
\begin{table}
    \centering
    \caption{Global fit parameter prior function type and prior limits for \Tstar\ b.}
     \label{tab:priorsb}
    \begin{tabular}{p{0.12\textwidth} l l}
    \toprule
    \textbf{Parameter} &\textbf{Prior} &\textbf{Value} \\
    \hline
    \textbf{\TStar\ b} &  \\
    Period (days) &$\mathcal{N}$(3.1837961, 0.0000112) &Table~\ref{tab:planet_props}\\
    Transit ephemeris (TBJD) &$\mathcal{U}$(2384.25, 2384.33)  &Table~\ref{tab:planet_props}\\
    log(\rpl) &$\mathcal{N}$(-3.549\S, 1)  & Table~\ref{tab:planet_props} (\rpl)\\
    Impact parameter &$\mathcal{U}$(0, 1)  &Table~\ref{tab:planet_props}\\
    Eccentricity & \citet{exoplanet:kipping13_2}, $\mathcal{B}$($e$, 0.867, 3.03)  &Table~\ref{tab:planet_props}\\
    Argument of periastron (rad) & $\mathcal{U}$($-\pi,\pi$)  &Table~\ref{tab:planet_props} ($^{\circ}$)\\
    \bottomrule
    \end{tabular}
    \begin{tablenotes}
    \item \textbf{Numbers in brackets represent:}
    \item (mean $\mu$, standard deviation $\sigma$) for normal distribution $\mathcal{N}$($\mu$,$\sigma$)
    \item (lower limit \textit{x}, upper limit \textit{y}) for uniform distribution $\mathcal{U}$(\textit{x},\textit{y})
    \item \S\ Equivalent to 0.5$\times$log($\delta$)+log(\rstar), $\delta$ represents transit depth (based on ExoFOP catalog values)
    \item Distributions for eccentricity \textit{e} are built into the \texttt{exoplanet} package and based on \citet{exoplanet:kipping13_2} which includes the Beta distribution $\mathcal{B}$($e,a,b$) (exponential $e$, shape parameter $a$, shape parameter $b$)
    \end{tablenotes}
\end{table}
\end{center}


\begin{center}
\begin{table}
    \centering
    \caption{Global fit parameter prior function type and prior limits for \tess\ photometric data.}
    \label{tab:priorsphot}
    \begin{tabularx}{0.5\textwidth}{ l l l }
    \toprule
    \textbf{Parameter} &\textbf{Prior}  &\textbf{Value} \\
    \hline
    \textbf{Sector 1} & \\
    Mean &$\mathcal{N}$(0, 1)   & -0.00001 $\pm$ 0.00009 \\
    log(\textit{s}2) &$\mathcal{N}$(-15.536$^*$ , 0)  & -15.781 $\pm$ 0.039 \\
    log(\textit{w}0) &$\mathcal{N}$(0, 0.1)     & 0.104 $\pm$ 0.095 \\
    log(\textit{Sw}4) &$\mathcal{N}$(-15.536$^*$ , 0)   & -15.574 $\pm$ 0.096 \\
    \hline
    \textbf{Sector 12} & \\
    Mean &$\mathcal{N}$(0, 1) & -0.00001 $\pm$ 0.00008 \\
    log(\textit{s}2) &$\mathcal{N}$(-15.702$^*$ , 0.1) & -15.822 $\pm$ 0.043 \\
    log(\textit{w}0) &$\mathcal{N}$(0, 0.1) & 0.132 $\pm$ 0.100 \\
    log(\textit{Sw}4) &$\mathcal{N}$(-15.702$^*$ , 0.1)   & -15.757 $\pm$ 0.100 \\
    \hline
    \textbf{Sector 13} & \\
    Mean &$\mathcal{N}$(0, 1) & 0.00004 $\pm$ 0.00014 \\
    log(\textit{s}2) &$\mathcal{N}$(-15.183$^*$ , 0.1) & -15.609 $\pm$ 0.040 \\
    log(\textit{w}0) &$\mathcal{N}$(0, 0.1) & 0.019 $\pm$ 0.088 \\
    log(\textit{Sw}4) &$\mathcal{N}$(15.183$^*$ , 0.1)   & -15.162 $\pm$ 0.095 \\
    \hline
    \textbf{Sector 27} & \\
    Mean &$\mathcal{N}$(0, 1) & 0.00007 $\pm$ 0.00033 \\
    log(\textit{s}2) &$\mathcal{N}$(-13.932$^*$ , 0.1) & -14.357 $\pm$ 0.027 \\
    log(\textit{w}0) &$\mathcal{N}$(0, 0.1) & 0.056 $\pm$ 0.094 \\
    log(\textit{Sw}4) &$\mathcal{N}$(-13.932$^*$ , 0.1)   & -13.411 $\pm$ 0.109 \\
    \hline
    \textbf{Sector 28} & \\
    Mean &$\mathcal{N}$(0, 1) & 0.00007 $\pm$ 0.00015 \\
    log(\textit{s}2) &$\mathcal{N}$(-14.578$^*$ , 0.1) & -14.646 $\pm$ 0.025 \\
    log(\textit{w}0) &$\mathcal{N}$(0, 0.1) & 0.139 $\pm$ 0.103 \\
    log(\textit{Sw}4) &$\mathcal{N}$(-14.578$^*$ , 0.1)   & -14.646 $\pm$ 0.098 \\
    \hline
    \textbf{Sector 39} & \\
    Mean &$\mathcal{N}$(0, 1) & 0.00002 $\pm$ 0.00012 \\
    log(\textit{s}2) &$\mathcal{N}$(-14.656$^*$ , 0.1) & -14.743 $\pm$ 0.022 \\
    log(\textit{w}0) &$\mathcal{N}$(0, 0.1) & 0.185 $\pm$ 0.104 \\
    log(\textit{Sw}4) &$\mathcal{N}$(-14.656$^*$ , 0.1)   & -14.761 $\pm$ 0.099 \\
    \bottomrule
    \end{tabularx}
    \begin{tablenotes}
    \item \textbf{Numbers in brackets represent:}
    \item (mean $\mu$, standard deviation $\sigma$) for normal distribution $\mathcal{N}$($\mu$,$\sigma$)
    \item \textbf{Prior values:}
    \item $^*$ Equivalent to the log of the variance of the \TESS\ flux from the corresponding sector
    \end{tablenotes}
\end{table}
\end{center}


\begin{center}
\begin{table}
    \centering
    \caption{Global fit parameter prior function type and prior limits for \harps\ radial velocity data.}
    \label{tab:priorsrv}
    \begin{tabularx}{0.5\textwidth}{ l l l }
    \toprule
    \textbf{Parameter} &\textbf{Prior}  &\textbf{Value} \\
    \hline
    \textit{K} (\ms) &$\mathcal{U}$(0, 10) &Table~\ref{tab:planet_props} \\
    Amplitude &$\mathcal{C}$(5) & 6.321 $\pm$ 2.657 \\
    \textit{l}\textsubscript{\textit{E}} &$\mathcal{T}$(20.53, 20, 20) & 34.982 $\pm$ 11.285 \\
    \textit{l}\textsubscript{\textit{P}} &$\mathcal{T}$(0.1, 10, 0, 1)  & 0.640 $\pm$ 0.209 \\
    \harps\ offset  & $\mathcal{N}$(9102.921$^{\dagger}$, 10) & 9103.035 $\pm$ 3.808  \\
    log(Jitter\textsubscript{\harps})  & $\mathcal{N}$(1.293$^{\ddagger}$, 5)  & -4.528 $\pm$ 2.708   \\
    \bottomrule
    \end{tabularx}
    \begin{tablenotes}
    \item \textbf{Numbers in brackets represent:}
    \item (lower limit \textit{x}, upper limit \textit{y}) for uniform distribution $\mathcal{U}$(\textit{x},\textit{y})
    \item (scale parameter $\beta$) for half-Cauchy distribution $\mathcal{C}$($\beta$)
    \item (mean $\mu$, standard deviation $\sigma$, lower limit \textit{x}, upper limit \textit{y}) for truncated normal distribution $\mathcal{T}$($\mu$,$\sigma$,\textit{x},\textit{y})
    \item (mean $\mu$, standard deviation $\sigma$) for normal distribution $\mathcal{N}$($\mu$,$\sigma$)
    \item \textbf{Prior values:}
    \item $^{\dagger}$ Equivalent to the mean of the \harps\ radial velocity
    \item $^{\ddagger}$ Equivalent to 2 times the log of the minimum \harps\ radial velocity error
    \end{tablenotes}
\end{table}
\end{center}


\begin{figure}
    \centering
    \includegraphics[width=0.5\textwidth]{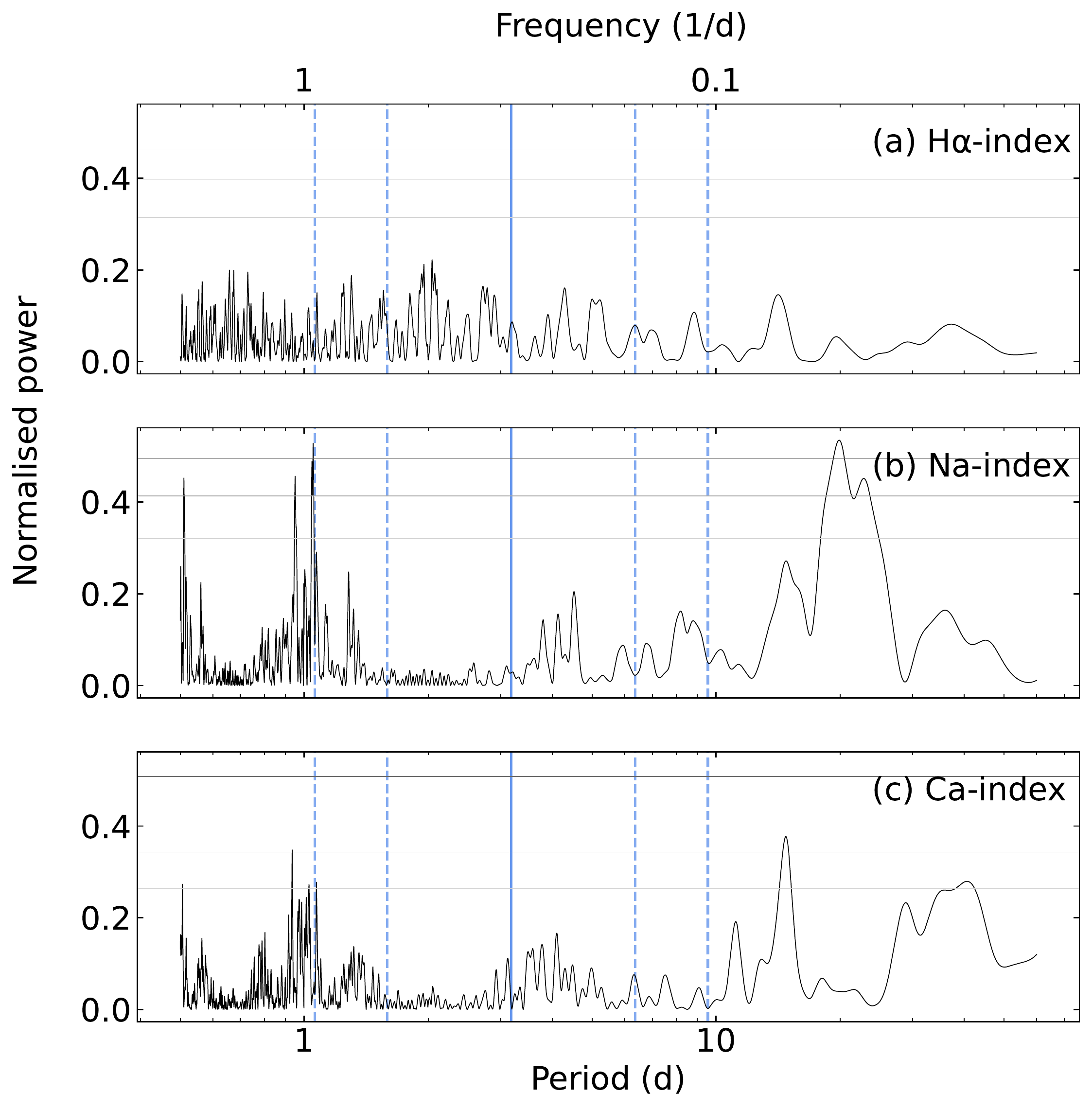}
    \caption{Periodograms for the \HARPS\ radial velocity data of the values presented in Table~\ref{tab:harpsappendix}. The orbital period of \Tplanetb\ is marked with a solid vertical blue line, alongside two period aliases either side of this value marked as dashed vertical blue lines. The False Alarm Probabilities (FAP) are marked as the horizontal grey lines at 0.1, 1 and 10\% from top to bottom. \textbf{Panels 1-3}: Periodograms for the H$\alpha$-index, Na-index and Ca-index.}
    \label{fig:harpsappendix}
\end{figure}

\begin{center}
\begin{table*}
    \centering
    \caption{Additional \HARPS\ activity indicator data for \Tstar. This table is available in its entirety online.}
    \label{tab:harpsappendix}
    \begin{tabular}{c c c c c c}  
    \toprule
    \textbf{H$\alpha$-index} & \textbf{H$\alpha$-index error} & \textbf{Na-index} & \textbf{Na-index error} & \textbf{Ca-index} & \textbf{Ca-index error} \\
    \hline
    0.003024 & 0.000001 & 0.264718 & 0.001861 & 0.121843 & 0.004656 \\
    0.002718 & 0.000001 & 0.260553 & 0.001679 & 0.132003 & 0.004192 \\
    0.002366 & 0.000001 & 0.263481 & 0.001440  & 0.132657 & 0.003269 \\
    0.002739 & 0.000001 & 0.260150  & 0.001688 & 0.121485 & 0.004012 \\
    0.002411 & 0.000001 & 0.261340  & 0.001464 & 0.139120  & 0.003539 \\
    ... & ...   & ...   & ...   & ...   & ... \\
\bottomrule
    \end{tabular}
\end{table*}
\end{center}


\begin{figure*}
    \centering
    \includegraphics[width=\textwidth]{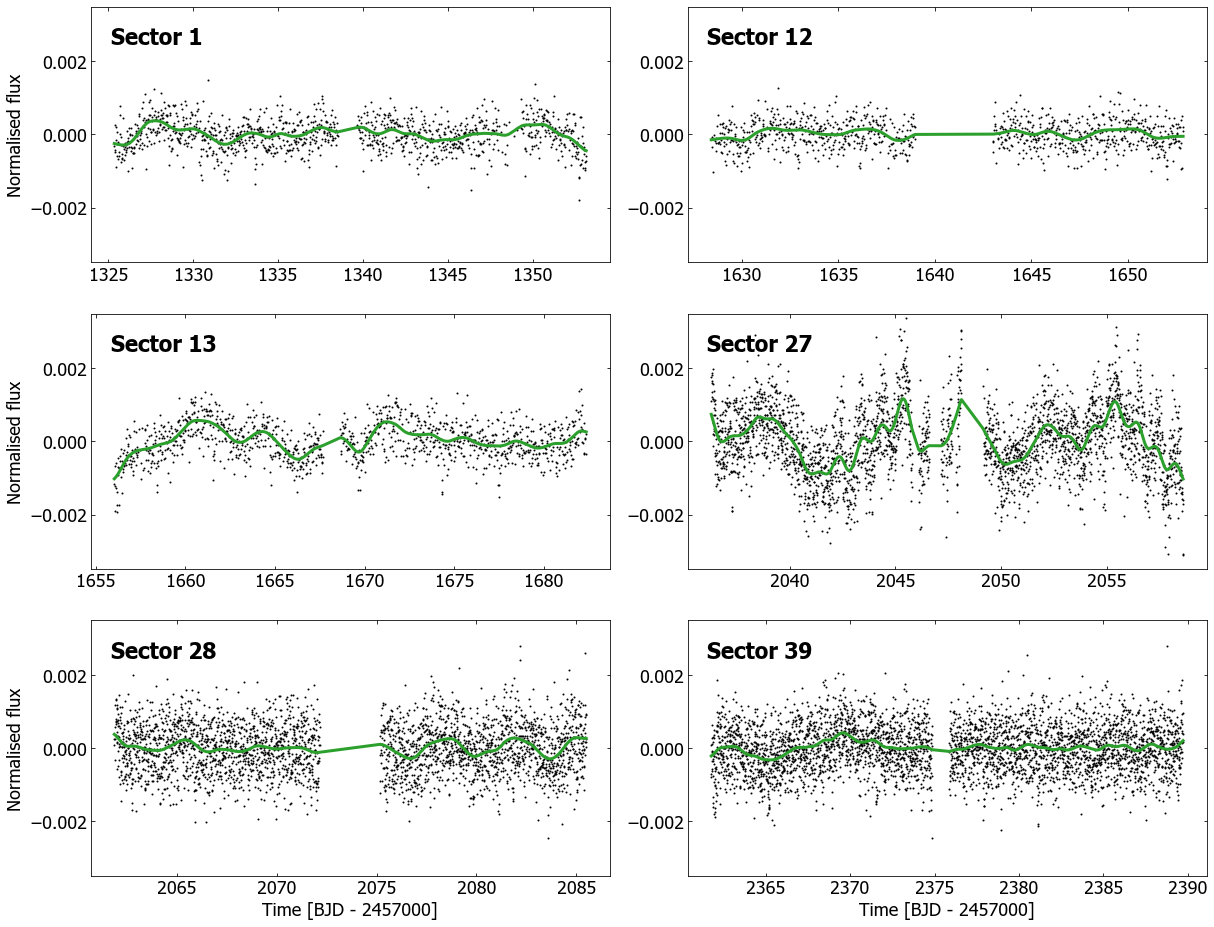}
    \caption{\tess\ PDCSAP lightcurves from labelled sectors, overplotted with the GP models in green.}
    \label{fig:gpmodels}
\end{figure*}

\begin{table}
    \caption{All TTVs from \tess.}
    \label{tab:ttvs}
\begin{tabular}{cccc}
\toprule
\textbf{Transit no.} & \textbf{TTV (mins)} & \textbf{Transit no.} & \textbf{TTV (mins)} \\
\hline
0   & $-5.4^{+8.9}_{-5.6}$   & 223 & $-4.7^{+5.7}_{-3.6}$    \\
1   & $-3.7^{+3.5}_{-8.4}$   & 224 & $3.0^{+1.0}_{-5.7}$     \\
2   & $15.9^{+1.8}_{-15.9}$  & 225 & $-10.3^{+20.3}_{-24.7}$ \\
3   & $4.2^{+1.8}_{-13.2}$   & 227 & $2.9^{+6.6}_{-3.0}$     \\
4   & $-9.5^{+9.4}_{6.86.8}$ & 228 & $-5.5^{+5.5}_{-1.3}$    \\
5   & $-1.9^{+7.6}_{-3.0}$   & 229 & $-29.4^{+29.2}_{-8.8}$  \\
6   & $0.9^{+10.2}_{-1.5}$   & 231 & $-5.0^{+11.3}_{-1.2}$   \\
7   & $2.6^{+8.1}_{-2.8}$    & 232 & $-1.0^{+4.7}_{-1.2}$    \\
8   & $-5.0^{+13.9}_{-1.1}$  & 233 & $4.7^{+0.3}_{-10.1}$    \\
95  & $23.9^{+14.3}_{-23.6}$ & 235 & $-0.5^{+2.5}_{-2.5}$    \\
96  & $-1.8^{+3.0}_{-4.4}$   & 236 & $2.3^{+14.9}_{-2.3}$    \\
97  & $-1.4^{+1.6}_{-4.6}$   & 237 & $-3.7^{+3.6}_{-2.8}$    \\
100 & $-7.7^{+10.2}_{-1.1}$  & 238 & $-5.3^{+6.4}_{-6.4}$    \\
101 & $11.4^{+2.5}_{-21.9}$  & 325 & $19.4^{+0.1}_{-20.1}$   \\
102 & $7.2^{+9.6}_{-7.1}$    & 326 & $-1.2^{+2.2}_{-7.6}$    \\
103 & $-8.1^{+12.7}_{-1.6}$  & 327 & $-1.4^{+1.4}_{-9.2}$    \\
104 & $8.1^{+6.1}_{-7.9}$    & 328 & $6.0^{+3.6}_{-6.2}$     \\
105 & $4.5^{+2.4}_{-13.5}$   & 329 & $-4.0^{+4.0}_{-11.9}$   \\
107 & $-9.7^{+9.6}_{-17.7}$  & 330 & $-15.7^{+15.4}_{-3.8}$  \\
108 & $3.1^{+1.8}_{-6.6}$    & 331 & $22.5^{+3.7}_{-22.3}$   \\
109 & $-4.3^{+8.7}_{-0.2}$   & 332 & $-2.5^{+10.4}_{-1.8}$   \\
110 & $10.5^{+1.3}_{-10.5}$  & 333 & $-9.0^{+10.4}_{-0.4}$   \\ \bottomrule
\end{tabular}
\end{table}


\section{Author affiliations} \label{sec:affiliations}


$^{1}$ Department of Physics, University of Warwick, Gibbet Hill Road, Coventry CV4 7AL, UK \\
$^{2}$ Centre for Exoplanets and Habitability, University of Warwick, Gibbet Hill Road, Coventry CV4 7AL, UK \\
$^{3}$ Instituto de Astrof\'isica e Ci\^encias do Espa\c{c}o, Universidade do Porto, CAUP, Rua das Estrelas, 4150-762 Porto, Portugal \\ 
$^{4}$ Departamento de F\'isica e Astronomia, Faculdade de Ci\^encias, Universidade do Porto, Rua do Campo Alegre, 4169-007 Porto, Portugal \\
$^{5}$ Physikalisches Institut, University of Bern, NCCR PlanetS, CSH, Gesellschaftsstrasse 6, 3012 Bern, Switzerland \\
$^{6}$ Center for Astrophysics \textbar\ Harvard \& Smithsonian, 60 Garden St., Cambridge, MA 02138, USA \\
$^{7}$ Observatoire de l’Universit\'e de Gen\`eve, Chemin Pegasi 51, 1290 Versoix \\
$^{8}$ INAF-Osservatorio Astrofisico di Torino, Via Osservatorio 20, I-10025 Pino Torinese, Italy \\
$^{9}$ IPAC-NASA Exoplanet Science Institute, 770 S. Wilson Avenue, Pasadena, CA 91106, USA \\
$^{10}$ Department of Astrophysical Sciences, Princeton University, Princeton, NJ 08544, USA\\
$^{11}$ International Center for Advanced Studies (ICAS) and ICIFI (CONICET), ECyT-UNSAM, Campus Miguelete, 25 de Mayo y Francia, (1650) Buenos Aires, Argentina \\ 
$^{12}$ Department of Astronomy and Tsinghua Centre for Astrophysics, Tsinghua University, Beijing 100084, China \\
$^{13}$ SUPA Physics and Astronomy, University of St. Andrews, Fife, KY16 9SS Scotland, UK \\ 
$^{14}$ Aix Marseille Univ, CNRS, CNES, LAM, Marseille, France \\ 
$^{15}$  NASA Ames Research Center, Moffett Field, CA 94035 \\
$^{16}$ Department of Physics \& Astronomy, Swarthmore College, Swarthmore PA 19081, USA \\
$^{17}$ Department of Physics and Astronomy, University of Louisville, Louisville, KY 40292, USA \\
$^{18}$ NASA Goddard Space Flight Center, 8800 Greenbelt Rd, Greenbelt, MD 20771, USA \\
$^{19}$ Department of Physics and Kavli Institute for Astrophysics and Space Research, Massachusetts Institute of Technology, Cambridge, MA 02139, USA \\
$^{20}$  Centro de Astrobiolog\'ia (CAB, CSIC-INTA), Depto. de Astrof\'sica, ESAC campus, 28692, Villanueva de la Ca\~nada (Madrid), Spain \\
$^{21}$ European Southern Observatory, Karl-Schwarzschild-Stra{\ss}e 2, 85748 Garching bei M{\"u}nchen, Germany \\
$^{22}$ Department of Earth, Atmospheric, and Planetary Sciences, Massachusetts Institute of Technology, Cambridge, MA 02139, USA \\
$^{23}$ Department of Aeronautics and Astronautics, Massachusetts Institute of Technology, Cambridge, MA 02139, USA \\ 
$^{24}$ Facultad de Ciencias Astron\'omicas y Geof\'isicas, Universidad Nacional de La Plata, Paseo del Bosque s/n, (B1900) Buenos Aires, Argentina \\
$^{25}$ Hazelwood Observatory, Australia \\
$^{26}$ SETI Institute, Mountain View, CA 94043, USA \\
$^{27}$ Department of Physics, Engineering and Astronomy, Stephen F. Austin State University, 1936 North St, Nacogdoches, TX 75962, USA \\


\bsp	
\label{lastpage}
\end{document}